\documentstyle[aps,prb,epsf,times]{revtex}
\begin{document}
\draft

\title{Pair Excitations, Collective Modes and Gauge Invariance in the
  BCS -- Bose-Einstein Crossover Scenario}
  \author{Ioan Kosztin, Qijin Chen, Ying-Jer Kao, and K. Levin}
  \address{The James Franck Institute, The University of Chicago, 5640
    South Ellis Avenue, Chicago, Illinois 60637}
\date{\today}
\maketitle
\begin{abstract}
  In this paper we study the BCS Bose Einstein condensation (BEC)
  crossover scenario within the superconducting state, using a T-matrix
  approach which yields the ground state proposed by Leggett.  Here we
  extend this ground state analysis to finite temperatures $T$ and
  interpret the resulting physics. We find two types of bosonic-like
  excitations of the system: long lived, incoherent pair excitations and
  collective modes of the superconducting order parameter, which have
  different dynamics. Using a gauge invariant formalism, this paper
  addresses their contrasting behavior as a function of $T$ and
  superconducting coupling constant $g$.  At a more physical level, our
  paper emphasizes how, at finite $T$, BCS-BEC approaches introduce an
  important parameter $\Delta^2_{pg} = \Delta^2 - \Delta_{sc}^2$ into
  the description of superconductivity.  This parameter is governed by
  the pair excitations and is associated with particle-hole asymmetry
  effects which are important for sufficiently large $g$. In the
  fermionic regime, $\Delta_{pg}^2$ represents the difference between
  the square of the excitation gap $\Delta^2$ and that of the
  superconducting order parameter $\Delta_{sc}^2$.  The parameter
  $\Delta_{pg}^2$, which is necessarily zero in the BCS (mean field)
  limit increases monotonically with the strength of the attractive
  interaction $g$.  It follows that there is a significant physical
  distinction between this BCS-BEC crossover approach (in which $g$ is
  the essential variable which determines $\Delta_{pg}$) and the widely
  discussed (Coulomb-modulated) phase fluctuation scenario in which the
  plasma frequency is the tuning parameter.  Finally, we emphasize that
  in the strong coupling limit, there are important differences between
  the composite bosons which arise in crossover theories, and the usual
  bosons of the (interacting) Bose liquid. Because of constraints
  imposed on the fermionic excitation gap and chemical potential, in
  crossover theories, the fermionic degrees of freedom can never be
  fully removed from consideration.
\end{abstract} 


\section{Introduction}
\label{sec:intro}

The observation of an excitation gap above $T_c$ (called the
``pseudogap'') in the underdoped cuprate superconductors has been the
focus of much current research. Presumably an understanding of this
state will help unravel the formal machinery, if not the attractive
pairing mechanism behind high temperature superconductivity. It is now
widely believed that this state is associated with the underlying
superconducting phase, in large part because the ($d$-wave) symmetry of
the pseudogap is found \cite{arpesanl,arpesstanford} to be the same as
that of the excitation gap and order parameter in the $T<T_c $ state.
Among viable candidates for the origin of the pseudogap state are phase
fluctuation scenarios\cite{emery95-434,franz98-14572}, $d$-wave nodal
excitation mechanisms \cite{lee97-4111} and a BCS Bose-Einstein
``crossover picture''
\cite{Randeria,Micnas,janko97-11407,uemura97-194,kosztin98-R5936,chen98-4708,maly99-1354,tchernyshyov97-3372,ranninger96-11961}.

The goal of the present paper is to discuss the last of these, the crossover
scenario within the superconducting state. Our work is directed towards the
fundamental issues of the crossover problem, with lesser emphasis on the
physics of the cuprates. We present a generalized overview based on a finite
temperature T matrix formulation.  Our aim is to provide a useful
understanding and to extend the physics of the well characterized ground
state\cite{Leggett}. In the process we establish a clear distinction between
incoherent, finite center of mass momentum, pair excitations and order
parameter fluctuations (i.e., collective modes), and their respective
dynamics.  We formulate a gauge invariant description of the electrodynamic
response with an emphasis on particle-hole asymmetry which is necessarily
very important.  Although we introduce a generalized T matrix approach,
special attention will be paid to one particular version which has been
extensively discussed in our previous work
\cite{janko97-11407,kosztin98-R5936,chen98-4708,maly99-1354,chen99-7083}.

In the BCS Bose-Einstein condensation (BEC) crossover approach, it is
presumed that there is a smooth evolution, with increasing attractive
coupling constant $g$, from BCS superconductivity, in which strongly
overlapping Cooper pairs form and Bose condense at precisely the same
temperature $T_c$, to a quasi-ideal Bose gas state (to be characterized in
more detail below) in which tightly bound fermion pairs (composite bosons)
form at temperatures much higher than their Bose condensation temperature
$T_c$. In this latter case there is an excitation gap (pseudogap) for
\textit{fermionic} excitations well above $T_c$.  These ideas date back to
the late sixties when Eagles\cite{Eagles} first drew attention to the
possibility of smoothly interpolating between the BCS and Bose-Einstein
ground state descriptions of superconductivity.  This was followed by a well
known paper by Leggett\cite{Leggett} who presented an interpolation scheme
for the ground state based on a variational wave-function, in which the
chemical potential $\mu$ was self consistently varied (with increasing $g$)
from $E_F$ to large negative values.  Nozieres and Schmitt-Rink\cite{NSR},
hereafter referred to as NSR, generalized these previous approaches by
presenting a crossover theory for computing $T_c$.  Although this leading
order theory was not fully self consistent\cite{Serene}, nevertheless, much
of the essential physics pertaining to finite $T$ was summarized in
Ref.~\onlinecite{NSR}.

With the discovery of the short coherence length cuprates, several
groups noted the relevance of this body of theoretical work.  Indeed,
this recognition was made well in advance of a community-wide
appreciation of pseudogap phenomena, which phenomena have only served to
re-enforce interest in these crossover schemes. Randeria and
co-workers\cite{Varenna} were among the first to apply the NSR approach
to the cuprates.  Micnas \textit{et al}\cite{micnas90-113} presented
detailed studies of the attractive Hubbard model with varying on-site
Coulomb interaction $U$, and Uemura \cite{uemura97-194} noted, on the
basis of unusual correlations deduced from $\mu$SR experiments, that the
cuprates exhibited aspects of bosonic character, as might be expected in
a crossover theory.  Our own group\cite{chen98-4708} has also addressed
cuprate issues in the past year using the formalism of the present
paper.

Attempts to go beyond the NSR scheme at finite $T$ are relatively more
recent and almost exclusively restricted to two dimensional systems.
Numerical simulations \cite{trivedi95-312,singer97-955} on the attractive
Hubbard model, along with numerical \cite{engelbrecht98-13406,deisz98} and
analytical \cite{tchernyshyov97-3372} studies of the so-called FLEX scheme
\cite{engelbrecht98-13406} have provided some insights.  This diagrammatic
FLEX approach should be contrasted with an alternative (called the ``pairing
approximation") which we
\cite{janko97-11407,kosztin98-R5936,chen98-4708,maly99-1354,chen99-7083}
have introduced into the literature and which is based on earlier work by
Kadanoff and Martin\cite{Kadanoff}, and extended by Patton\cite{Patton}.  In
contrast to the FLEX scheme this latter approach precisely yields BCS theory
in the small $g$ limit.  What is more important in distinguishing our work
from that of others, however, is our direct
focus\cite{kosztin98-R5936,chen98-4708} on \textit{crossover effects within
  the superconducting state}. This is the topic of the present paper, as
well, and necessarily requires studies of systems in higher than two
dimensions, where $T_c$ is non-zero.  The ultimately decisive factor in
determining whether these crossover theories or any other alternatives are
appropriate for the cuprates, may well come from the predicted behavior
below $T_c$.  This fact has also been emphasized by Deutscher
\cite{deutscher99-410}.

\section{Overview of BCS-BEC Crossover Theories}
\label{sec:BCS-BEC_overview}
\subsection{Physical Picture at $T=0$: Previous Work}
\label{ssec:phys_pict1}

All crossover theories are based on an underlying ``generic'' Hamiltonian
describing the attractive interaction between
fermions with opposite spin orientation (for spin singlet pairing)

\begin{eqnarray}
{\cal H} & = & \sum_{{\bf k}\sigma} \epsilon_{\bf k}
c^{\dag}_{{\bf k}\sigma} c^{\ }_{{\bf k}\sigma}
\nonumber \\
& & + \sum_{\bf k k' q} V_{\bf k, k'} 
c^{\dag}_{{\bf k}+{\bf q}/2\uparrow} 
c^{\dag}_{-{\bf k}+{\bf q}/2\downarrow} 
c^{\ }_{-{\bf k'}+{\bf q}/2\downarrow} 
c^{\ }_{{\bf k'}+{\bf q}/2\uparrow},
\label{eq:2}
\end{eqnarray}
where $c^\dagger_{{\bf k}\sigma}$ creates a particle in the momentum
state ${\bf k}$ with spin $\sigma $, and $\epsilon_{\mathbf{k}}$ is the
energy dispersion measured from the chemical potential $\mu$ (we take
$\hbar=k_B=1$). For the jellium case $\epsilon_{\mathbf{k}}={\bf
  k}^2/2m-\mu$, while for the lattice case we consider an anisotropic
tight-binding model with $\epsilon_{\mathbf{k}}=2t_{\parallel}(2-\cos
k_x - \cos k_y) + 2t_{\perp}(1-\cos k_z)-\mu$, where $t_{\parallel}$
($t_{\perp}$) is the hopping integral for the in-plane (out-of-plane)
motion, and we set the lattice constant $a=1$. Here we assume a
separable pairing interaction $V_{\bf k,k'} = g \varphi_{\bf
  k}\varphi_{\bf k'}$, where $g=-|g|$ is the coupling strength and the
momentum dependence of the function $\varphi_{\mathbf{k}}$, reflects the
pairing anisotropy, and its form depends on the particular model under
consideration. For the jellium case we take\cite{NSR}
$\varphi_{\mathbf{k}} = (1+k^2/k_0^2)^{-1/2}$, where $1/k_0$ gives the
range of the interaction and represents a soft cutoff in momentum space
for the pairing interaction. In the lattice case, $\varphi_{\mathbf{k}}
= 1$ for $s$-wave pairing symmetry, and $\varphi_{\mathbf{k}} = \cos k_x -
\cos k_y$ for $d$-wave symmetry. To simplify the notation, until Sec.~V,
when the effects of $\varphi_{\mathbf{k}}$ become relevant, we choose
to write our equations with $\varphi_{\mathbf{k}} = 1$.  

It is important to stress the key assumptions of these crossover theories:
(i) only two body fermionic interactions are included. (ii) In calculations
of \textit{equilibrium} properties (as distinct from studies of the
collective mode of the superconducting order parameter), repulsive Coulomb
interactions between fermions are absorbed into the effective pairing
interaction $V_{\mathbf{k},\mathbf{k}'}$. These Coulomb effects are presumed
to be weak enough so that the attractive interactions driving
superconductivity dominate, and it is assumed that Coulomb interactions do
\textit{not} occur between fermion pairs or composite bosons.

Using this Hamiltonian, Leggett\cite{Leggett} found that the BCS ground
state wavefunction
\begin{equation}
  |\Psi\rangle = \Pi_{\bf k} \left( u_{\bf k} + v_{\bf k} c^\dagger_{{\bf k}
      \uparrow}c^\dagger_{{\bf {-k}} \downarrow}\right)|\text{vac}\rangle
\end{equation}
is appropriate to \textit{both} the weakly and strongly interacting
limits, provided $v^2_{\bf k}=\frac{1}{2}(1-\epsilon_{\bf k}/E_{\bf k}),
u^2_{\bf k}=\frac{1}{2}(1+\epsilon_{\bf k}/E_{\bf k})$, where $E_{\bf k}
= \sqrt{\epsilon^2_{\bf k}+\Delta^2}$.  Moreover, the explicit parameter
$\Delta$ and the implicit parameter $\mu$ follow from the following two
self consistent conditions

\begin{equation}
  g^{-1} + \sum_{\bf k}\frac{1}{2E_{\bf k}}\, = 0\;,
\label{Thouless-crit}
\end{equation}
and 

\begin{equation}
  n = 2\sum_{\bf k} v^2_{\bf k} \;,
  \label{number}
\end{equation}
where $n$ is the electron number density.

In the weak coupling limit, the above equations yield the usual BCS state.
By contrast, in the strong coupling (i.e., $g\rightarrow\infty$) limit, for
the case of jellium, the system corresponds to a Bose condensation of
nonoverlapping and \textit{non-interacting} tightly bound pairs of fermions,
which resemble to diatomic molecules. This is an ``essentially ideal Bose
gas''\cite{NSR}.  Indeed, it can be shown\cite{Leggett} that in the strong
coupling limit, Eq.~(3) can be recast in the form of the Schr\"oedinger
equation, written in momentum space, for an \textit{isolated} diatomic
molecule, consisting of the two fermions.  The strong coupling limit was
argued to be slightly more complicated for a lattice of fermions\cite{NSR}
which, at moderate densities, is to be distinguished from jellium. Here a
canonical transformation can be used to partially integrate out the fermions
and the Pauli principle then leads to hard core repulsions between composite
bosons.  Nevertheless, even for a lattice, this scheme is viewed as
reasonable\cite{NSR} so that the BCS wave function assumption, or
equivalently Eqs.~(\ref{Thouless-crit}) and (\ref{number}), represent an
effective mean field approximation to the solution of the hard core
composite boson problem.
This essentially ideal Bose gas treatment of the ground state in the
strong coupling limit is to be contrasted with the behavior found in the
collective mode spectrum, where calculations lead to results similar to
those for a weakly interacting Bose gas. These calculations suggest the
presence of an \textit{effective} boson-boson interaction in this limit.

Within crossover theories, the $T=0$ behavior of the Anderson Bogoliubov
(AB) mode has been studied in Refs.~\onlinecite{belkhir92-5087},
\onlinecite{cote93-10404} and \onlinecite{Micnas}. In the long wavelength
(small $\mathbf{q}$) limit the dispersion relation of the sound-like AB mode
is linear: $\omega=cq$, where $c$ is the AB mode velocity. In weak coupling,
the usual BCS limit is obtained for the AB mode velocity $c = v_F/\sqrt{3}$.
In strong coupling, $c$ reflects an effective boson-boson interaction
\cite{engelbrecht97-15153} which arises from the Pauli principle.  We
caution that even though the phrase boson-boson interaction is frequently
used here and in the literature, in the composite boson system (within the
crossover scenario), these interactions should always be viewed as indirect
and associated with the underlying fermionic interactions.

In summary, the $T=0$ description of the composite boson problem (in the
strong coupling limit) represents a cross between ideal and non-ideal
Bose-Einstein physics.  Whereas, an incomplete condensation at zero
temperature is required in order to obtain superfluidity in a ``true"
Bose system\cite{fetter}, in the crossover scenario, the condensation is
always complete (at $T=0$), but, nevertheless the phase mode and
superconductivity exist.

\subsection{Physical Picture at Finite $T< T_c$: Present Work}
\label{ssec:phys_pict2}

In this section we discuss our general results and physical picture for the
finite $T$ crossover problem, at an intuitive level. These results are then
re-derived in more detail in Sec.~III using a microscopic T-matrix approach.
As the temperature is increased above $T=0$, and the coupling $g$ becomes
sufficiently strong, two important effects ensue \cite{kosztin98-R5936}.
(i) The excitation gap is no longer the same as (the amplitude of) the
superconducting order parameter.  (ii) Incoherent pair excitations
with non-zero center of mass momentum can be thermally excited.  Indeed,
points (i) and (ii) are inter-related. The first of these is anticipated as
one leaves the BCS regime, where even above $T_c$, there is expected to be a
normal state gap for fermionic excitations; this gap is associated with
meta-stable (intermediate $g$) \cite{janko97-11407} or stable (large $g$)
pairs.  It is natural to presume, as we have found,\cite{kosztin98-R5936}
that these effects persist below $T_c$.  Indeed, if there is an excitation
gap at $T_c$, in any second order superconducting transition, this gap must
necessarily be different from the superconducting order parameter, at least,
at and slightly below $T_c$.

The second of these two points was noted in Ref.~\onlinecite{NSR}.  These
pair excitations are related to the ``particle" excitation branch of the
interacting Bose system, which (although similar to collective phase mode
branch at small wave-vectors), represents a distinct dynamical
mode\cite{fetter}.  In the composite boson problem the pair excitations and
related pair propagator, called ${\cal T} ( i\Omega , {\bf q })$, play an
essential role.  In the limit of non-zero $Q\equiv(i\Omega,{\bf q})$, we
refer to this propagator as ${\cal T}_{pg}(Q)$, where the subscript $pg$
derives from ``pseudogap''. For small $\Omega$ and ${\bf q }$, it may be
approximated by

\begin{equation}
\label{eq:4a}
{\cal T}_{pg} (i\Omega,{\bf q}) \approx Z/
(i\Omega - \Omega_{\bf q} + i \Gamma_{\bf q}  +\mu_{pair})\;, 
\end{equation}
where $Z$ is the usual renormalization factor, $i\Omega$ is a bosonic
Matsubara frequency, $\Omega_{\bf q}$ is the dispersion of the finite
momentum pair excitations (with $\Omega_{{\bf q}=0}=0$), $\Gamma_{\bf
  q}^{-1}$ is the pair excitation lifetime, and $\mu_{pair}$ is the
associated effective pair ``chemical potential'' with its value set by
the well known (ideal gas) BEC condition

\begin{equation}
\label{eq:4b}
\mu_{pair} = 0\;, \qquad \text{for}\quad T \le T_c \;.
\end{equation}
It is this function ${\cal T }_{pg}$, which will be shown in Sec.~III, to
determine the \textit{fermionic} excitation gap $\Delta (T)$ and chemical
potential $\mu$.  Moreover, the fermionic excitation gap is related, in
turn, to the condition that $\mu_{pair} = 0 $ at and below $T_c$. Quite
generally, the presence of incoherent pair excitations blocks the available
states for fermions and, thereby, affects the fermion excitation gap. These
effects, which occur both above and below $T_c$, are directly related to
particle-hole asymmetry which appears as the system crosses out of the BCS
regime.

It should be stressed that within the BCS-BEC crossover picture, the form of
${\cal T}$ is highly circumscribed so as to produce the ground state
equations, Eqs.~(\ref{Thouless-crit}) and (\ref{number}).  Through these
equations the fermionic degrees of freedom play an important role, even at
very strong coupling.  This constraint can then be used to deduce the
function $\Omega_{\mathbf{q}}$.  We find that the pairing approximation
(discussed in Sec.~III below) produces this $T=0$ description of the
fermionic degrees of freedom, and that, as a consequence, for sufficiently
small ${\bf q}$, the pair excitation dispersion is given by

\begin{equation}
\label{eq:4c}
 \Omega_{\bf q} = {\bf q}^2 / 2 M_{\text{pair}} 
\end{equation}
where the pair mass $M_{\text{pair}}$ is dependent on $g$,
temperature, density,
lattice structure and other materials properties.  This dispersion is found
to be a consequence of other T-matrix schemes, as well, and takes a similar
form above $T_c$, although there the pairs are not as long lived.  Once the
pair propagator is characterized, $T_c$ can be obtained either when
approached from above or below.

We may now quantify the first point (i) listed above. The deviation between
the excitation gap and the order parameter is related to the number of
thermally excited finite momentum pair excitations. We define the difference
between the excitation gap $\Delta$ and order parameter $\Delta_{sc}$ as

\begin{equation}
 \label{eq:1}
 \Delta_{pg}^2 =   \Delta ^ 2  - \Delta_{sc}^2\;,
\end{equation}
where throughout this paper, $\Delta_{sc}$ is taken to be real.  The number
of (incoherent) pair excitations is given in terms of the pair propagator
${\cal T}_{pg}$ 

\begin{equation}
\label{eq:4}
\Delta_{pg}^2 = -\sum_Q{\cal T}_{pg}(Q) = -\sum_{\bf
  q}\int^{\infty}_{-\infty} \frac{{\mbox{d}}\Omega}{\pi}\, b(\Omega)
\,{\text{Im}}\, {\cal T}_{pg} (\Omega,{\bf q})\;,
\end{equation}
where $b(\Omega)$ is the Bose function and ${\cal T}_{pg}(\Omega,
\mathbf{q})$ is the analytically continued ($i\Omega\rightarrow
\Omega+i0^+$) pair propagator.
In the regime of intermediate coupling, where pseudogap effects are
apparent, these bosonic excitations act in concert with fermionic
excitations\cite{chen98-4708}. This is what one expects if there is to be a
smooth interpolation between the BCS and BEC limits.

The behavior of $\Delta_{pg}$, $\Delta$ and $\Delta_{sc}$ is
schematically plotted in Fig.~1 in the three different regimes: weak
(BCS), intermediate, and strong coupling (nearly BEC) regimes.  Below
$T_c$ these plots are based on detailed numerical
calculations\cite{kosztin98-R5936}, whereas above $T_c$, where the
computations are more difficult\cite{maly99-1354}, on a simple
extrapolation procedure.\cite{AboveTcExtrapolation}

It may be noted that Fig.~1 has a direct analogue in the (Coulomb modulated)
phase fluctuation scenario\cite{emery95-434}. The three panels (from top to
bottom) would then correspond to progressively decreasing the size of the
plasma frequency $\omega_p$. In this phase fluctuation scenario the tuning
parameter is $\omega_p$, whereas in the crossover scenario the parameter $g$
sets the scale for the size of $\Delta_{pg}^2$.  It should, thus, be clear
that Coulomb-modulated phase fluctuations are not the only way to create an
excitation gap which appears above the transition temperature.

It should, finally, be stressed that these incoherent, finite momentum pair
excitations, which enter into $\Delta_{pg}$ via Eq.~(\ref{eq:4}), are
irrelevant in the BCS limit, in accord with Eq.~(\ref{eq:1}) and Fig.~1(a).
In that limit the ``quasi-particle'' assumption implicit in
Eq.~(\ref{eq:4a}) is invalid and because of both the lack of particle-hole
asymmetry and the large damping $\Gamma_{\mathbf{q}}$, the pair excitation
spectrum overlaps the particle-particle continuum.  One can quantify the
reliability of this key approximation.  In Fig.~2, using our numerical
scheme\cite{chen98-4708}, we plot the value $\Lambda$ of the wave-vector
$|{\bf q}|$ at which the pair dispersion intersects the continuum states, as
a function of coupling $g$. This corresponds to a measure of the Landau
damping of the pair propagator. The shaded region indicates where the
incoherent finite momentum pairs represent ill-defined excitations.  Outside
this shaded region, the assumptions implicit in Eq.~(\ref{eq:4a}) should be
valid.  Related calculations show that $\Delta_{pg} (T_c) / \Delta_{sc} (0)
$ is arbitrarily small in the weak coupling limit, so that even if we apply
Eq.~(\ref{eq:4a}) directly in this limit, pseudogap effects are negligible
in the BCS regime.

\subsection{The Strong Coupling Limit: ``True" versus  Composite Bosons }

A first important distinction can be found, between true and composite
bosons, at the level of the Leggett ground state.  As noted earlier, already
at $T=0$ there appears to be a mix of quasi-ideal and interacting Bose gas
character to the strong coupling limit. The gap equation
[Eq.~(\ref{Thouless-crit})] is associated \cite{Leggett} with
``noninteracting diatomic molecules", whereas, the collective mode
spectrum \cite{belkhir92-5087,Micnas} reflects an effective boson-boson
interaction which relates to the Pauli statistics of the constituent
fermions.  This is revealed most clearly in jellium models where the AB
sound velocity remains finite at infinite $g$, with an asymptote associated
with the residual interactions.  Some insight into the origin of these
boson-boson effects can be found in Refs.~\onlinecite{haussmann93-291} and
\onlinecite{haussmann94-12975}.
Indeed, one should be cautious in the use of the phrase effective
boson-boson repulsion which, for large $g$, derives entirely from Fermi
statistical effects. Unlike in the interacting Bose system\cite{fetter},
where boson-boson interactions need to be separately included in the boson
propagator, here the physics associated with the Pauli principle is already
accounted for and should not be fed back again to renormalize the dispersion
of the composite bosons.

A second important difference arises from the fact that this
superconducting ground state corresponds to one in which there is full
condensation so that, as in the BCS phase, the condensate fraction $n_0 =
n$. By contrast, in a Bose superfluid there is always a depletion of the
condensate at $T=0$, caused by the existence of the boson-boson repulsion.

As a final important difference, we note that the behavior of the pair
propagator ${\cal T}$, which must necessarily be consistent with
Eqs.~(\ref{Thouless-crit}) and (\ref{number} ), is highly circumscribed
and rather different from what one might deduce based on the Bogoliubov
model for a Bose liquid\cite{fetter}.  The fermionic degrees of freedom
can never be fully ``integrated out''. The fermionic excitation gap
$\Delta$ and the pair chemical potential $\mu_{pair}=0$ are, moreover,
closely inter-related via, e.g., Eq.~(\ref{eq:t4}) below. These effects
have no natural counterpart in the Bose liquid (where the fermionic
excitation gap is of no consequence).  It is, thus, not surprising that
in $T_c$ calculations, which are associated with a divergence in ${\cal
  T }$, a variety of different groups
\cite{chen99-7083,haussmann93-291,haussmann94-12975} find that the
strong coupling limit is characterized by a quasi-ideal BEC result

\begin{equation}
\label{eq:16}
  T_c = \frac{3.31}{M_{pair}}\left(\frac{n}{2}\right)^{2/3}
\end{equation}
where $M_{pair}$ is the same pair mass as that which appears in
$\Omega_{\bf q} = {\bf q}^2/M_{pair} $ of Eq.~(\ref{eq:4c}).  Thus,
from the perspective of $T_c$ and of the \textit{temperature dependent} gap
equations, \textit{the bosons are ``free''}, except for the renormalized
mass.  Related theories have reached rather similar conclusions.
\cite{NSR,Randeria,Drechsler} These conclusions owe their origin to the
underlying mean field structure of BCS theory, which is the starting point
of the crossover scenario. At a more microscopic level, they appear to be
associated with general T-matrix approaches, as is discussed below.

\section{Self-consistent T-matrix Approximations}
\label{sec:SCTMA}
\subsection{General Results for the Superconducting Phase}
\label{ssec:below_Tc}

In this section we present the self consistency conditions and related
gap equations associated with the superconducting state within the broad
class of T-matrix-based crossover theories.  An important goal of this
discussion is to show that Eqs.~(\ref{eq:4a})-(\ref{eq:4}) are rather
general consequences of these schemes when applied below $T_c$.  In
these non-perturbative approaches the two-particle Green's function is
expressed in terms of a T-matrix (or pair propagator), ${\cal T } (Q) $,
which is determined self consistently in terms of the single particle
Green's function.  One solves, in effect, three coupled equations for
the self energy, T-matrix and chemical potential (via the number
equation):
\begin{mathletters}
\label{eq:t1}
\begin{equation}
  \label{eq:t1a}
  \Sigma(K) = G_o^{-1}(K)-G^{-1}(K) =
  \sum_Q {\cal T}(Q)\,\tilde{G}(Q-K)\;,
\end{equation}
\begin{equation}
  \label{eq:t1b}
  g \;=\; [1+g\,\chi(Q)]\,{\cal T}(Q)\;,
\end{equation}
\begin{equation}
  \label{eq:t1c}
  n \;=\; 2\sum_K G(K)\;.
\end{equation}
\end{mathletters}
Here and in what follows we use four-vector notation: $Q \equiv
(i\Omega,{\bf q})$, $\sum_Q\equiv T\sum_{i\Omega}\sum_{\mathbf{q}}$,
etc.  The choice of the functions $\tilde{G}$ and pair susceptibility
$\chi$ varies from one approximation to another.  Here we address two
different schemes which introduce self consistency at a level beyond the
lowest order T-matrix approximation used by NSR,\cite{NSR} so that
our discussion focuses on T-matrix schemes where dressed Green's
functions enter into the self consistency requirements.  (a) The first
approach, known as the FLEX (or `$GG$') approximation
\cite{tchernyshyov97-3372,haussmann94-12975,engelbrecht98-13406} takes
$\tilde{G}=G$ with $\chi=\chi^{FLEX}$ where we define

\begin{equation}
\chi^{FLEX} (P) = \sum_K G(K)G(P-K)
\end{equation}
with the corresponding self energy
\begin{equation}
\Sigma^{FLEX} (K) = \sum_P {\cal T } (P) G (P-K)
\end{equation}
In addition, we study the (b) pairing approximation
\cite{kosztin98-R5936,chen98-4708} (or `$GG_o$'scheme), which sets
$\tilde{G}=G_o$ and $\chi=\chi^{pair}$ with
\begin{equation}
\chi^{pair} (P) = \sum_K G(K)G_o(P-K)
\end{equation}
with the corresponding self energy
\begin{equation}
\Sigma^{pair} (K) = \sum_P {\cal T } (P) G_o (P-K)
\end{equation}

We know of no other literature on the FLEX scheme within the superconducting
state (appropriate to a fully three dimensional system).  By contrast, above
$T_c$ there is considerable literature on the behavior of ${\cal T}$ and
$T_c$, which, in the FLEX scheme, have been calculated numerically
\cite{haussmann94-12975} and analytically \cite{tchernyshyov97-3372} by
solution of Eqs.~(\ref{eq:t1}).
As the temperature is lowered, this T-matrix develops a maximum around
$Q=0$ and the transition temperature $T_c$ to the broken symmetry phase
is signaled by a pole given by the condition
\begin{equation}
  \label{eq:t2}
  g\, {\cal T}^{-1}(Q=0;T_c) = 1+g\chi(Q=0;T_c,\Delta) = 0\;.
\end{equation}

The counterpart of this analysis is considerably more complicated below
$T_c$. The procedure, which we summarize in this section, is an
approximation which is chosen to satisfy the following four criteria:
(i) It leads to the same transition temperature when approached from
below as from above.  (ii) It leads directly to a \textit{physical}
interpretation of the BCS-BEC crossover scheme as discussed in the
previous section and illustrated in Fig.~1.  An important, and third
criterion which we view as an additional check on the approximations
used is that (iii) one should recover the BCS scheme in weak coupling
for all $T\le T_c$ and, finally, (iv) we should recover the Leggett ground
state at $T=0$.

An important premise underlying these crossover schemes is that the
particle-particle dominates the particle-hole channel. Without such an
assumption one would be forced to evaluate a full $2\times 2$ Nambu
matrix Green's function along with a $4\times 4$ T-matrix.  A complete
self-consistent solution of the resulting set of equations is
prohibitively difficult. With the assumption that the particle-particle
channel is dominant, one argues that formally the set of
Eqs.~(\ref{eq:t1}) remains valid below $T_c$, provided that the T-matrix
acquire a singular delta-function component, which describes the ($Q=0$)
Cooper pair condensate in equilibrium, so that
\begin{mathletters}
  \label{eq:t3}
\begin{equation}
  \label{eq:t3a}
  {\cal T}(Q) = {\cal T}_{sc}(Q) + {\cal T}_{pg}(Q)\;,
\end{equation}
where 
\begin{equation}
  \label{eq:t3b}
  {\cal T}_{sc}(Q) = -\frac{\Delta_{sc}^2}{T}\delta(Q)\;, \quad
  \text{and}\quad {\cal T}_{pg}(Q) = \frac{g}{1+g\chi(Q)}\;.
\end{equation}
\end{mathletters}
This form for ${\cal T}$ guarantees that the regular part ${\cal
  T}_{pg}$ of the T-matrix remains finite for any non-zero $Q=
(i\Omega,{\bf q})$.  Inserting Eqs.~(\ref{eq:t3}) into
Eq.~(\ref{eq:t1b}) and treating separately the delta function
contribution and regular terms, one arrives at the following quite
general gap equation
\begin{equation}
  \label{eq:t4}
  1+g\chi(Q=0;T,\Delta) = 0\;, \qquad T\le T_c\;.
\end{equation}
In our self consistent scheme, the critical temperature $T_c$ can be
obtained from Eq.~(\ref{eq:t4}) by setting $\Delta_{sc}$, (which is
implicitly contained in $\Delta$), to zero. This result coincides precisely
with the Thouless criterion of Eq.~(\ref{eq:t2}), which is obtained by
approaching the transition from above.  As a result of the form of
Eq.~(\ref{eq:t4}), the regular part of the T-matrix ${\cal T}_{pg}(Q)$
diverges as $Q \rightarrow 0$; it can be written in the form

\begin{equation}
\label{eq:t5}
{\cal T }_{pg} ( i\Omega , q ) = Z / ( i\Omega - \Omega_q + i\Gamma_q)\;,
\qquad\quad T\le T_c\;,
\end{equation}
in accord with Eq.~(\ref{eq:4a}). Here $\Gamma_{\bf q }\rightarrow 0$ as
${\bf q }\rightarrow 0$.
Equation~(\ref{eq:t4}) also leads directly to a microscopic derivation
of Eq.~(\ref{eq:1}).  The self energy of Eq.~(\ref{eq:t1a}) may be
decomposed into two terms:

\begin{equation}
  \label{eq:sigma-sc-pg}
  \Sigma(K) \;=\; \Sigma_{sc}(K) + \Sigma_{pg}(K)\;,
\end{equation}
where the term associated with the condensate contribution, called
${\cal T }_{sc} $ is

\begin{equation}
  \label{eq:sigma-sc}
  \Sigma_{sc}(K) = -\Delta_{sc}^2 \tilde{G}(-K)\;.
\end{equation}
In evaluating the pseudogap contribution to $\Sigma$, as a consequence
of Eq.~(\ref{eq:t4}), the main contribution to the $Q$ sum comes from
the small $Q$ region, so that the integral may be approximated by

\begin{equation}
  \label{eq:sigma-pg}
  \Sigma_{pg}(K) \approx \tilde{G}(-K)\;\sum_Q{\cal T}_{pg}(Q) =
  -\Delta_{pg}^2 \tilde{G}(-K)\;. 
\end{equation}
In this way the total self energy is
\begin{equation}
  \label{eq:sigma}
  \Sigma(K) \approx -\Delta^2 \tilde{G}(-K)\;, \quad 
\Delta\equiv\sqrt{\Delta_{sc}^2+\Delta_{pg}^2}\;.
\end{equation}
Here the total excitation gap is to be associated with the parameter
$\Delta$, and we see that Eqs.~(\ref{eq:sigma-pg}) and (\ref{eq:sigma})
naturally lead back to the definition of $\Delta_{pg}$ which appears in
Eq.~(\ref{eq:1}).

There is another way of relating $\Delta_{pg}^2$ to the fluctuations of
the pairing field, which leads to a more precise interpretation of
Eq.~(\ref{eq:1}).  We write the pairing field $\hat{\Delta}_{{\bf
    q}}(t)\equiv |g|\sum_{{\bf k}} c_{-{\bf k}+{\bf q}/2\,\downarrow}(t)
c_{{\bf k}+{\bf q}/2\,\uparrow}(t)$, about its mean field value
$\langle|\Delta|\rangle\equiv|\langle\hat{\Delta}_{{\bf q}=0}(0)\rangle|
= \Delta_{sc}$. We define $\langle|\Delta|^2\rangle\equiv
\sum_{Q}\langle\hat{\Delta}_Q\hat{\Delta}^{\dag}_Q\rangle =
g^2\sum_{K,K'} \sum_Q C_2(K,K';Q)$, where $C_2(K,K';Q)$ is the proper
two-particle correlation function, which in general is not factorizable
in the variables $K$ and $K'$. After some algebra one arrives at
\begin{equation}
  \label{eq:t6}
  \langle|\Delta|^2\rangle = -\sum_Q {\cal T}(Q)[g\,\chi(Q)]^2\;.
\end{equation}
Now, inserting in Eq.~(\ref{eq:t6}) the expression for the T-matrix
given by Eqs.~(\ref{eq:t3}), we obtain
\begin{equation}
  \label{eq:t7}
  \langle|\Delta|^2\rangle = \Delta_{sc}^2 - \sum_Q {\cal
    T}_{pg}(Q)[g\,\chi(Q)]^2 \;.
\end{equation}
Since ${\cal T}_{pg}$ is highly peaked about $Q=0$, the pair
susceptibility $\chi(Q)$ on the right hand side in Eq.~(\ref{eq:t7})
can be approximated by its $Q=0$ value. We again make use of the gap
equation (\ref{eq:t4}) to write $\sum_Q {\cal T}_{pg}(Q)[g\,\chi(Q)]^2
\approx -\sum_Q {\cal T}_{pg}(Q) = \Delta_{pg}^2$.
Thus, quite generally 

\begin{equation}
\label{eq:t8}
  \Delta_{pg}^2 = - \sum_Q
  {\cal T }_{pg}(Q) \approx \langle|\Delta|^2\rangle -
  \langle|\Delta|\rangle^2 \;. 
\end{equation}

The above discussion is expected to apply to both the FLEX scheme and
the pairing approximation. Moreover, on the basis of the behavior above
$T_c$, there is no \textit{a priori} reason to select one approach over
the other.  However, the latter seems to be preferred if one imposes the
third and fourth criteria discussed above.

Indeed, the superconducting state is associated with the self energy of
Eq.~(\ref{eq:sigma-sc}).  It can be seen that $\Sigma_{sc}$ coincides
with the BCS self energy if we adopt the pairing approximation so that
$\tilde{G}=G_o$.  With this result the BCS gap equation follows from
Eq.~(\ref{eq:t4}):
\begin{equation}
  \label{eq:t6a}
  1+g\sum_Q G(K)G_o(-K) = 1 + g\sum_K 
\frac{\Delta_{sc}^2}{\omega^2+E_{\bf k}^2} =0\;.
\end{equation}

The results of the pairing approximation, at all $g$, can be summarized
simply.  This scheme leads to the following equations for the three
unknowns $\Delta$ , $\mu$, and $\Delta_{pg}$

\begin{mathletters}
\begin{equation}
g^{-1} + \sum_{\bf k}\frac{1-2f(E_{\bf k})}{2E_{\bf 
k}}\, = 0\;,
\label{Thouless-crit2}
\end{equation}
where $E_{\bf k} = \sqrt{\epsilon^2_{\bf k}+\Delta ^2 }$.  This equation
must be solved self consistently with
\begin{equation}
  n = 2\sum_{\bf k}\left[v^2_{\bf k} + 
    \frac{\epsilon_{\bf k}}{E_{\bf k}}\,f(E_{\bf k})\right] \;,
  \label{number2}
\end{equation}
where $v^2_{\bf k}=\frac{1}{2}(1-\epsilon_{\bf k}/E_{\bf k})$.

Finally, the decomposition of $\Delta$ into $\Delta_{sc}$ and $\Delta_{pg}$
requires the solution of a third equation, namely Eq.~(\ref{eq:4}), which we
repeat here for completeness

\begin{equation}
\label{eq:5d}
\Delta_{pg}^2 = -\sum_Q{\cal T}_{pg}(Q) = -\sum_{\bf
  q}\int^{\infty}_{-\infty} \frac{{\mbox{d}}\Omega}{\pi}\, b(\Omega)
\,{\text{Im}}\, {\cal T}_{pg} (\Omega,{\bf q})\;.
\end{equation}
\label{eq:5}
\end{mathletters}

An additional check on the validity of the approximation scheme relates
to criterion (iv), that is, the nature of the ground state. It follows
from Eq.~(\ref{eq:5d}) that, as a result of the Bose function $b(\Omega)$,
quite generally
\begin{equation}
  \lim_{T\rightarrow 0} \Delta_{pg} = 0 \;,
\end{equation}
as is consistent with a physical picture in which $\Delta_{pg}$ is
associated with classical fluctuations.  When this limit is used within
the pairing approximation, the ground state which results from
Eqs.~(\ref{Thouless-crit2}) and (\ref{number2}) is the same as that
proposed by Leggett [see Eqs.~(\ref{Thouless-crit}) and (\ref{number})].


\section{Electromagnetic Response and Collective Modes of a 
  Superconductor: Beyond BCS Theory }
\label{sec:gauge}

The purpose of this section is to study the gauge invariant (linear)
response of a superconductor to an external electromagnetic (EM) field, and
obtain the associated collective mode spectrum.  Our discussion is generally
relevant to complex situations such as those appropriate to the BCS-BEC
crossover scenario.  An important ingredient of this discussion is
establishing the role of particle-hole asymmetry.
It should be noted that there are fairly extensive discussions in the
literature on the behavior of collective modes within the $T=0$ crossover
scenario \cite{cote93-10404,Micnas,belkhir92-5087}. Here we review a
slightly different formulation\cite{kulik81-591,zha95-6602} which introduces
a matrix extension of the Kubo formalism of the normal state.  We find that
this approach is more directly amenable to extension to finite $T$, where
the pair fluctuation diagrams need to be incorporated.

The definition of the collective modes of a superconductor must be made with
some precision. We refer to the underlying Goldstone boson of the charged or
uncharged superconductor as the ``AB mode'', after Anderson
\cite{anderson58-1900} and Bogoliubov \cite{bogoliubov58-794}. According to
our specific definition, this AB mode appears as a pole structure in the
gauge invariant formulations of the electrodynamic response functions, for
example, in the density-density correlation function. Early work by
Prange\cite{prange63-2495} referred to this as the ``ghost mode'' of the
neutral system, since this term is not directly affected by the long range
Coulomb interaction.  By contrast, the normal modes of the charged or
uncharged superconductor, which we shall call the ``collective modes'',
involve a coupling between the density, phase, and for the BCS-BEC case,
amplitude degrees of freedom.  For these, one needs to incorporate a many
body theoretic treatment of the particle-hole channel as well.  In crossover
theories this channel is not as well characterized as is the
particle-particle channel.

\subsection{Gauge Invariant EM Response Kernel}
\label{ssec:EM-K}

In the presence of a weak externally applied EM field, with four-vector
potential $A^{\mu} = (\phi, {\bf A})$, the four-current density $J^{\mu}
= (\rho, {\bf J})$ is given by
\begin{equation}
  \label{eq:em1}
  J^{\mu}(Q) = K^{\mu\nu}(Q) A_{\nu}(Q)\;,
\end{equation}
where, $Q\equiv q^{\mu}=(\omega,{\bf q})$ is a four-momentum, and
$K^{\mu\nu}$ is the EM response kernel which can be written as
\begin{equation}
  \label{eq:em4}
  K^{\mu\nu}(Q) = K_0^{\mu\nu}(Q) + \delta{K}^{\mu\nu}(Q)\;.
\end{equation} 
Here
\begin{equation}
  \label{eq:em2}
  K_0^{\mu\nu}(\omega,{\bf q}) = P^{\mu\nu}(\omega,{\bf q}) + 
\frac{ne^2}{m}
  g^{\mu\nu}(1-g^{\mu 0})
\end{equation}
is the usual Kubo expression for the electromagnetic response.  We define
the current-current correlation function $P^{\mu\nu}(\tau,{\bf q}) =
-i\theta(\tau)\langle[j^{\mu}(\tau,{\bf q}),j^{\nu}(0,-{\bf q})]\rangle$.
In the above equation, $g^{\mu\nu}$ is the contravariant diagonal metric
tensor, with diagonal elements $(1,-1,-1,-1)$, and $n$, $e$ and $m$ are the
particle density, charge and mass, respectively.

The presence of $\delta{K}^{\mu\nu}$ in Eq.~(\ref{eq:em4}) is due to the
perturbation of the superconducting order parameter by the EM field,
i.e., to the excitation of the \textit{collective modes} of
$\Delta_{sc}$. This term is required to satisfy charge conservation
$q_{\mu}J^{\mu}=0$, which requires that
\begin{mathletters}
\label{eq:em3}
  \begin{equation}
    \label{eq:em3a}
    q_{\mu}K^{\mu\nu}(Q) = 0\;.
  \end{equation}
Moreover,  gauge invariance yields
\begin{equation}
  \label{eq:em3b}
  K^{\mu\nu}(Q) q_{\nu}  = 0\;,
\end{equation}
\end{mathletters}
Note that, since $K^{\mu\nu}(-Q) = K^{\nu\mu}(Q)$, the two constraints
Eqs.~(\ref{eq:em3}) are in fact equivalent.

The incorporation of gauge invariance into a general microscopic theory may
be implemented in several ways.  Here we do so via a general matrix linear
response approach \cite{kulik81-591} in which the perturbation of the
condensate is included as additional contributions $\Delta_1+i\Delta_2$ to
the applied external field.  These contributions are self consistently
obtained (by using the gap equation) and then eliminated from the final
expression for $K^{\mu\nu}$.  We now implement this procedure.  Let
$\eta_{1,2}$ denote the change in the expectation value of the pairing field
$\hat{\eta}_{1,2}$ corresponding to $\Delta_{1,2}$.  For the case of an
$s$-wave pairing interaction $g<0$, the self-consistency condition
$\Delta_{1,2}=-g\eta_{1,2}$ leads to the following equations
\begin{mathletters}
  \label{eq:em9}
  \begin{equation}
    \label{eq:em9a}
    J^{\mu} = K^{\mu\nu}A_{\nu} = K_0^{\mu\nu}A_{\nu} + R^{\mu
      1}\Delta_1 + R^{\mu 2}\Delta_2\;,
  \end{equation}
  \begin{equation}
    \label{eq:em9b}
    \eta_1 = -\frac{\Delta_1}{g} = R^{1\nu}A_{\nu} + Q_{11}\Delta_1 +
    Q_{12}\Delta_2\;,
  \end{equation}
  \begin{equation}
    \label{eq:em9c}
    \eta_2 = -\frac{\Delta_2}{g} = R^{2\nu}A_{\nu} + Q_{21}\Delta_1 +
    Q_{22}\Delta_2\;,
  \end{equation}
\end{mathletters}
where $R^{\mu i}(\tau,{\bf q})=-i\theta(\tau)\langle[j^{\mu}(\tau,{\bf q}),
\hat{\eta}_i(0,-{\bf q})]\rangle$, with $\mu=0,\ldots,3$, and $i=1,2$; and
$Q_{ij}(\tau,{\bf q})=-i\theta(\tau)\langle[\hat{\eta}_i(\tau,{\bf q}),
\hat{\eta}_j(0,-{\bf q})]\rangle$, with $i,j=1,2$.

Thus far, the important quantities $K_0^{\mu\nu}$, $R^{\mu i }$ and
$Q_{ij}$ are unknowns which contain the details of the appropriate
microscopic model.  We shall return to these later in
Sec.~\ref{sec:PG_EM_exp}. The last two of Eqs.~(\ref{eq:em9}) can be
used to express $\Delta_{1,2}$ in terms of $A_{\nu}$
\begin{mathletters}
  \label{eq:em10}
  \begin{equation}
    \label{eq:em10a}
    \Delta_1 = -\frac{\tilde{Q}_{22}R^{1\nu} -
      Q_{12}R^{2\nu}}{\tilde{Q}_{11}\tilde{Q}_{22} - Q_{12}Q_{21}} 
A_{\nu}\;,
  \end{equation}
  \begin{equation}
    \label{eq:em10b}
    \Delta_2 = -\frac{\tilde{Q}_{11}R^{2\nu} -
      Q_{21}R^{1\nu}}{\tilde{Q}_{11}\tilde{Q}_{22} - Q_{12}Q_{21}} 
A_{\nu}\;,
  \end{equation}
\end{mathletters}
where $\tilde{Q}_{ii} = 1/g+Q_{ii}$, with $i=1,2$. Finally, inserting
Eqs.~(\ref{eq:em10}) into Eq.~(\ref{eq:em9a}) one obtains

\begin{mathletters}
\label{eq:em11}
\begin{equation}
  \label{eq:em11a}
  K^{\mu\nu} = K_0^{\mu\nu} + \delta{K}^{\mu\nu}\;,
\end{equation}
with
\begin{equation}
  \label{eq:em11b}
\hspace*{-8ex}\delta{K}^{\mu\nu} = -\frac{\tilde{Q}_{11}R^{\mu 
2}R^{2\nu} +
  \tilde{Q}_{22}R^{\mu 1}R^{1\nu} - Q_{12}R^{\mu 1}R^{2\nu} - 
Q_{21}R^{\mu
  2}R^{1\nu}}{\tilde{Q}_{11}\tilde{Q}_{22} - Q_{12}Q_{21}} \;.
\end{equation}
\end{mathletters}

As can be seen from the above rather complicated equation, the
electromagnetic response of a superconductor involves many different
components of the generalized polarizability.  Moreover, in the form of
Eqs.~(\ref{eq:em11b}) it is not evident that the results are gauge
invariant. In order to demonstrate gauge invariance and reduce the
number of component polarizabilities, we first rewrite $K^{\mu\nu}$ in a
way which incorporates the effects of the amplitude contributions via a
renormalization of the relevant generalized polarizabilities, i.e.,
\begin{mathletters}
  \label{eq:em14}
  \begin{equation}
    \label{eq:em14a}
    K^{\mu\nu} = {K'}_0^{\mu\nu} + \delta{K'}^{\mu\nu}\;,
  \end{equation}
where
\begin{equation}
  \label{eq:em14b}
  {K'}_0^{\mu\nu} = K_0^{\mu\nu}-\frac{R^{\mu
      1}R^{1\nu}}{\tilde{Q}_{11}}\;,
\end{equation}
and
\begin{equation}
  \label{eq:em14c}
  {R'}^{\mu 2} = R^{\mu 2}-\frac{Q_{12}}{\tilde{Q}_{11}}R^{\mu 2}\;, 
\quad
  \tilde{Q}'_{22} = \tilde{Q}_{22} - 
\frac{Q_{12}Q_{21}}{\tilde{Q}_{11}}\;.
\end{equation}
\end{mathletters}
In this way we obtain a simpler expression for $\delta{K'}^{\mu\nu}$
\begin{equation}
  \quad \delta{K'}^{\mu\nu} = -\frac{R'^{\mu
      2}R'^{2\nu}}{\tilde{Q'}_{22}}\;.
\label{em:deltaK'}
\end{equation}

We now consider a particular (\textit{a priori} unknown) gauge
$A'^{\mu}$ in which the current density can be expressed as
$J^{\mu}={K'}_0^{\mu\nu}A'_{\nu}$.  The gauge transformation
\cite{klemm88-139} which connects the four-potential $A_{\mu}$ in an
arbitrary gauge with $A'_{\mu}$, i.e., $A'_{\mu}=A_{\mu}+i\chi q_{\mu}$,
must satisfy

\begin{equation}
  \label{eq:em5}
  J^{\mu} = K^{\mu\nu}A_{\nu} = {K'}_0^{\mu\nu}\left(A_{\nu}+i\chi
  q_{\nu}\right)\;.
\end{equation}
Now invoking charge conservation, one obtains
\begin{equation}
  \label{eq:em6}
  i\chi = -
\frac{q_{\mu}{K'}_0^{\mu\nu}A_{\nu}}{q_{\mu'}{K'}_0^{\mu'\nu'}q_{
\nu'}}\;, 
\end{equation}
and, therefore,

\begin{equation}
  \label{eq:em15}
  K^{\mu\nu} = {K'}_0^{\mu\nu} - 
\frac{\left({K'}_0^{\mu\nu'}q_{\nu'}\right)
\left(q_{\nu''}{K'}_0^{\nu''\nu}\right)}{q_{\mu'}{K'}_0^{\mu'\nu'}q_{
\nu'}}\;.
\end{equation}
The above equation satisfies two important requirements: it is manifestly
gauge invariant and, moreover, it has been reduced to a form that depends
principally on the four-current-current correlation functions.  (The word
``principally'' appears because in the absence of particle-hole symmetry,
there are effects associated with the order parameter amplitude
contributions which enter via Eq.~(\ref{eq:em14}) and add to the complexity
of the calculations).  Equation~(\ref{eq:em15}) should be directly compared
with Eq.~(\ref{eq:em11b}).  In order for the formulations to be consistent
and to explicitly keep track of the conservation laws (\ref{eq:em3}), the
following identities must be satisfied:

\begin{mathletters}
  \label{eq:em13}
  \begin{equation}
    \label{eq:em13a}
    \left(q_{\mu}{K'}_0^{\mu\nu}\right)\tilde{Q'}_{22} =
    \left(q_{\mu}{R'}^{\mu 2}\right) {R'}^{2\nu}\;, 
  \end{equation}
  \begin{equation}
    \label{eq:em13b}
    \left({K'}_0^{\mu\nu}q_{\nu}\right)\tilde{Q'}_{22} =
    {R'}^{\mu 2}\left({R'}^{2\nu}q_{\nu}\right) \;.
  \end{equation}
\end{mathletters}
These identities may be viewed as ``Ward identities'' for the
superconducting two particle correlation functions\cite{zha95-6602}.  Any
theory which adds additional self energy contributions to the BCS scheme
must obey these important equations.  We shall return to this issue in
Sec.~\ref{sec:PG_EM_exp}.

\subsection{The Goldstone Boson or AB Mode}
\label{ssec:AB_mode}

The EM response kernel [cf.~Eqs.~(\ref{eq:em14})-(\ref{eq:em15})] of a
superconductor contains a pole structure which is related to the underlying
Goldstone boson of the system.  Unlike the phase mode component of the
collective mode spectrum, this AB mode is independent of Coulomb effects
\cite{prange63-2495}. The dispersion of this amplitude renormalized AB mode
is given by

\begin{equation}
  \label{eq:em19}
  q_{\mu}{K'}^{\mu\nu}q_{\nu} = 0\;.
\end{equation}
For an isotropic system ${K'}_0^{\alpha\beta} = {K'}_0^{11}
\delta_{\alpha\beta}$, and Eq.~(\ref{eq:em19}) can be rewritten as

\begin{equation}
  \label{eq:em20}
  \omega^2 {K'}_0^{00} + {\bf q}^2 {K'}_0^{11} - 2\omega q_{\alpha}
  {K'}_0^{0\alpha} = 0\;,
\end{equation}
with $\alpha=1,2,3$, and in the last term on the LHS of Eq.~(\ref{eq:em20})
a summation over repeated Greek indices is assumed.
It might seem surprising that from an analysis which incorporates a
complicated matrix linear response approach, the dispersion of the AB
mode ultimately involves only the amplitude renormalized four-current
correlation functions, namely the density-density, current-current and
density-current correlation functions. This result is, nevertheless, a
consequence of gauge invariance.

At zero temperature ${K'}_0^{0\alpha}$ vanishes, and the sound-like AB mode
has the usual linear dispersion $\omega=\omega_{\bf q}=c|{\bf q}|$ with the
``sound velocity'' given by

\begin{equation}
  \label{eq:em22}
  c^2 = {K'}_0^{11}/{K'}_0^{00} \;.
\end{equation}
The equations in this section represent an important starting point for
our numerical analysis.
 
\subsection{General Collective Modes}

We may interpret the AB mode as a special type of collective mode which
is associated with $A_{\nu} = 0 $ in Eqs.~(\ref{eq:em10}). This mode
corresponds to free oscillations of $\Delta_{1,2}$ with a dispersion
$\omega= c q$ given by the solution to the equation

\begin{equation}
  \label{eq:em17}
  \text{det}|Q_{ij}| = \tilde{Q}_{11}\tilde{Q}_{22}-Q_{12}Q_{21} = 0\;.
\end{equation}
More generally, according to Eq.~(\ref{eq:em9a}) the collective modes of
the order parameter induce density and current oscillations. In the same
way as the pairing field couples to the mean field order parameter in
the particle-particle channel, the density operator $\hat{\rho}(Q)$
couples to the mean field $\delta\phi(Q)=V(Q)\delta\rho(Q)$, where
$V(Q)$ is an effective particle-hole interaction which may derive from
the pairing channel or, in a charged superconductor, from the Coulomb
interaction. Here $\delta\rho = \langle\hat{\rho}\rangle-\rho_0$ is the
expectation value of the charge density operator with respect to its
uniform, equilibrium value $\rho_0$.  Within our self-consistent linear
response theory the field $\delta\phi$ must be treated on an equal
footing with $\Delta_{1,2}$, and formally can be incorporated into the
linear response of the system by adding an extra term $K_0^{\mu
  0}\delta\phi$ to the right hand side of Eq.~(\ref{eq:em9a}). The other
two Eqs.~(\ref{eq:em9}) should be treated similarly. Note that, quite
generally, the effect of the ``external field'' $\delta\phi$ amounts to
replacing the scalar potential $A^0=\phi$ by $\bar{A^0} = \bar{\phi} =
\phi+\delta\phi$. In this way one arrives at the following set of three
linear, homogeneous equations for the unknowns $\delta\phi$, $\Delta_1$,
and $\Delta_2$

\begin{mathletters}
  \label{eq:em23}
  \begin{eqnarray}
    0 &=& R^{10}\delta\phi + \tilde{Q}_{11}\Delta_1 + Q_{12}\Delta_2\;,
    \label{eq:em23a} 
\\ 0 &=& R^{20}\delta\phi + Q_{21}\Delta_1 +
    \tilde{Q}_{22}\Delta_2\;, 
\label{eq:em23b} 
\\ \delta\rho =
    \frac{\delta\phi}{V} &=& K_0^{00}\delta\phi + R^{01}\Delta_1 +
    R^{02}\Delta_2 \;.  
\label{eq:em23c}
  \end{eqnarray}
\end{mathletters}
The dispersion of the collective modes of the system is given by the
condition that the above Eqs.~(\ref{eq:em23}) have a nontrivial 
solution

\begin{equation}
  \label{eq:em24}
  \left| 
    \begin{array}{ccc}
      Q_{11}+1/g & Q_{12} & R^{10} \\
      Q_{21} & Q_{22}+1/g & R^{20} \\
      R^{01} & R^{02} & K_0^{00} - 1/V
    \end{array}
  \right| = 0\;.
\end{equation}
In the case of particle-hole symmetry $Q_{12}=Q_{21}=R^{10}=R^{01}=0$
and, the amplitude mode decouples from the phase and density modes; the
latter two are, however, in general coupled.


\section{Effect of Pair Fluctuations on the Electromagnetic Response: 
  Some Examples}
\label{sec:PG_EM_exp}

Once dressed Green's functions $G$ enter into the calculational schemes, the
collective mode polarizabilities (e.g., $Q_{22}$) and the EM response tensor
$K_0^{\mu\nu}$ must necessarily include vertex corrections dictated by the
form of the self energy $\Sigma$, which depends on the T-matrix ${\cal T }$,
which, in turn depends on the form of $\chi$.  These vertex corrections are
associated with gauge invariance and with the constraints which are
summarized in Eqs.~(\ref{eq:em13}). It can be seen that these constraints
are even more complicated than the Ward identities of the normal state.
Indeed, it is relatively straightforward to introduce collective mode
effects into the electromagnetic response in a completely general fashion
which is required by gauge invariance.  This issue was discussed in
Sec.~\ref{sec:gauge} as well as extensively in the literature
\cite{JRS-sc,cote93-10404}. The difficulty is in the implementation.  In
this section we begin with a discussion of the $T=0$ behavior where the
incoherent pair excitation contributions to the self energy corrections and
vertex functions vanish. In this section, we shall keep the symmetry factor
$\varphi_{\mathbf{k}}$ explicitly.

\subsection{$T=0$ Behavior of the AB Mode  and Pair Susceptibility}

It is quite useful to first address the zero temperature results since there
it is relatively simple to compare the associated polarizabilities of the AB
mode with that of the pair susceptibility $\chi$.  In the presence of
particle hole symmetry this collective mode polarizability can be associated
with $Q_{22}$ which was first defined in Eq.~(\ref{eq:em9c}).  In the more
general case (which applies away from the BCS limit) $Q_{22}$ must be
replaced by a combination of phase and amplitude terms so that it is given
by $Q'_{22} = Q_{22} - Q_{12}Q_{21}/Q_{11}$.

We may readily evaluate these contributions in the ground state, where
$\Delta_{sc} = \Delta $. The polarizability $Q_{22}$ is given by 

\begin{equation}
Q_{22}(Q)=\frac{1}{2}\sum_P
\left[G(-P)G(P-Q)+G(P)G(Q-P)+F^\dagger(P)F^\dagger(P-Q)
+F(P)F(P-Q)\right] \varphi_{{\mathbf{p}}-{\mathbf{q}}/2}^2,
\end{equation}
where 
\begin{equation}
G(K) = G_o(K)/[1+ \Delta_{sc}^2\varphi_{\mathbf{k}}^2 G_o(-K)G_o(K)]\;,
\end{equation}
and

\begin{equation}
F (P) =  \Delta_{sc}\varphi_{\mathbf{p}} G(P) G_o(-P)\;.
\end{equation}

Now, it can be seen that the pair susceptibility $\chi$ in the pairing
approximation satisfies

\begin{equation}
  \sum_P [G(-P)G(P)+F(P)F(P)]\varphi_{\mathbf{p}}^2 = \sum_P
  G(P)G_o(-P)\varphi_{\mathbf{p}}^2 = \chi(0)
\end{equation}
and, moreover, $Q_{12}(0) = Q _{21} (0) = 0 $ so that
\begin{equation}
\frac{1}{g}+Q_{22}(0)=\frac{1}{g}[1+g\chi(0)] = 0 \;.
\end{equation}
In this way, the AB mode propagator is soft under the same conditions which
yield a soft pair excitation propagator ${\cal T }_{pg} = g/(1+g \chi )$,
and these conditions correspond to the gap equation
Eq.~(\ref{Thouless-crit}).  Moreover, it can be seen that $Q'_{22} ( Q ) =
Q'_{22}(-Q)$ so that, upon expanding around $Q=0$, one has
$Q_{22}(Q)=\alpha_{22}\Omega^2-\beta_{22}q^2$, $Q_{12}(Q) = -Q_{21}(Q) =
-i\Omega\alpha_{12}$, and $1/g+Q_{11}(Q)=1/g+\alpha_{11}$, where
\begin{eqnarray}
\alpha_{22}&=&\sum_{\mathbf{k}}
\frac{\varphi_{\mathbf{k}}^2}{8E_{\mathbf{k}}^3} \;,
\nonumber\\
\beta_{22}&=&\frac{1}{d}\sum_{\mathbf{k}}
  \frac{1}{8E_{\mathbf{k}}^3} \left[\varphi_{\mathbf{k}}^2
    (\vec{\nabla} \epsilon_{\mathbf{k}})^2 - \frac{1}{4} (\vec{\nabla}
    \epsilon_{\mathbf{k}}^2)\cdot
    (\vec{\nabla}\varphi_{\mathbf{k}}^2)\right] \;,
\nonumber\\
\alpha_{12}&=&\sum_{\mathbf{k}}
\frac{\epsilon_{\mathbf{k}}}{4E_{\mathbf{k}}^3} \varphi_{\mathbf{k}}^2 \;,
\nonumber\\
\alpha_{11}&=&\sum_{\mathbf{k}}\frac{\epsilon_{\mathbf{k}}^2}
{2E_{\mathbf{k}}^3} \varphi_{\mathbf{k}}^2 \;,
\label{eq:expansion}
\end{eqnarray}
where $d$ denotes the dimensionality of the system. Thus, one obtains

\begin{equation}
c^2=\frac{\beta_{22}} 
{\alpha_{22}+\frac{\alpha_{12}^2}{1/g+\alpha_{11}}}.
\end{equation}
At weak coupling in 3D, where one has particle-hole symmetry, $\alpha_{12}=
0$, the amplitude and the phase modes decouple. This leads to the well known
result $c=v_F/\sqrt{3}$. More generally, for arbitrary coupling strength
$g$, these equations yield results equivalent to those in the
literature\cite{belkhir92-5087,Micnas,cote93-10404}, as well as those
derived from the formalism of Sec.~IV.B.
Finally, it should be noted that, since both Eqs.~(\ref{eq:em15}) and
(\ref{em:deltaK'}) have the same poles, the condition $Q'_{22} (Q) = 0$
yields the same AB mode dispersion as that determined from
Eq.~(\ref{eq:em19}). This is a consequence of gauge invariance.

\subsection{AB Mode at Finite Temperatures}

We now turn to finite temperatures where there is essentially no prior
work on the collective mode behavior in the crossover scenario.  At the
level of BCS theory (and in the Leggett ground state) the extended
``Ward identities'' of Eqs.~(\ref{eq:em13}) can be explicitly shown to
be satisfied.  Presumably they are also obeyed in the presence of
impurities, as, for example, in the scheme of
Ref.~\onlinecite{kulik81-591}. However, in general, it is difficult to
go beyond these simple cases in computing all components of the matrix
response function.  Fortunately, the calculation of the AB mode is
somewhat simpler. It reduces to a solution of Eq.~(\ref{eq:em20}), which,
\textit{in the presence of particle-hole symmetry}, involves a
computation of only the electromagnetic response kernel: the
density-density, density-current and current-current correlation
functions.

It is the goal of this section to compute these three response functions
within the ``pairing approximation'' to the T-matrix.  Our work is based on
the normal state approach of Patton\cite{Patton} and the associated diagrams
are shown in Fig.~3.  Because full Green's functions $G$ appear in place of
$G_o$ (as indicated by the heavy lines) these diagrams are related to but
different from their counterparts studied by Aslamazov and Larkin and by
Maki and Thompson\cite{Patton}.  This diagram scheme forms the basis for
calculations published by our group\cite{kosztin98-R5936,chen99-7083} of the
penetration depth within the BCS-BEC crossover scheme.

Here we make one additional assumption. We treat the amplitude
renormalizations which appear in Eqs.~(\ref{eq:em14}), only
approximately since these contributions introduce a variety of
additional correlation functions, which must be calculated in a
consistent fashion, so as to satisfy Eqs.~(\ref{eq:em13}).  Because the
amplitude mode is gapped, at least at low $T$, we can approximate these
amplitude renormalizations by their $T=0$ counterparts, which are much
simpler to deduce.

The three electromagnetic correlation functions reduce to a calculation
of $P^{\mu\nu}$, which can be written as

\begin{equation}
  P^{\mu\nu}(Q)=2\sum_K \lambda^\mu (K,K-Q) G(K)G(K- Q)\Lambda^\nu (K,K-Q)
\label{P_mu_nu}\,,
\end{equation}
where 
$\lambda(K,K-Q)=(1, \frac{\partial \epsilon_{{\bf k}-{\bf q}/2}}
{\partial {\bf k}})$ and $\Lambda(K,K-Q)=\lambda(K,K-Q)+\delta
\Lambda_{sc}(K,K-Q) + \delta \Lambda_{pg}(K,K-Q)$ are the bare and full
vertices, respectively.

To evaluate the vertex function $\Lambda^{\mu} $ we decompose it into a
pseudogap contribution $\Lambda_{pg}$ and a superconducting contribution
$\Lambda_{sc}$. (The latter can be regarded as the Gor'kov ``$F$'' function
contribution, although we do not use that notation here).  The pseudogap
contribution comes from a sum of Maki-Thompson (MT) and Aslamazov-Larkin
(AL$_{1,2}$) type of diagrams [see Fig.~3(b)].  Since these vertex
corrections can be obtained from a proper vertex insertion to the self
energy, it follows that there is a cancellation between these various terms
which simplifies the algebra.  This cancellation is shown in more detail in
Appendix~\ref{ap:1}.
Following the analysis in this Appendix, the sum of both $pg$ and $sc$
contributions is given by
\begin{eqnarray}
\label{Vertex}
  \delta \Lambda^\mu (K, K-Q)&\approx& -(\Delta_{sc}^2-
\Delta_{pg}^2)
  \varphi_{\mathbf{k}} \varphi_{\mathbf{k-q}} G_0(-K) G_0(Q-K)
  \lambda^\mu(Q-K,-K)\nonumber\\ 
&& - \Delta_{pg}^2 G_0(-K)\frac{\partial
    \varphi^2_{{\mathbf{k}}-{\mathbf{q}}/2}}{\partial k_\mu} \,,
\end{eqnarray}
where use has been made of the fact that ${\cal T}_{pg}(Q)$ is highly peaked at
$Q=0$, and that $\Delta_{pg}^2\equiv - \sum_Q {\cal T}_{pg}(Q)$.

The AB mode dispersion involves the sum of three terms which enter into
Eqs.~(\ref{eq:em19}) and (\ref{eq:em20}). 
We next substitute Eqs.~(\ref{Vertex}) into Eq.~(\ref{P_mu_nu}). After
performing the Matsubara frequency summation, and analytically
continuing $i\Omega \rightarrow \Omega+i0^+$, we obtain for small $\Omega$
and $\textbf{q}$ 
\begin{eqnarray}
  q_\mu K_0^{\mu\nu}q_\nu &=& 
{\mathbf{q}}\cdot \left(\tensor{\frac{n}{m}}+\tensor{\bf P}\right) \cdot
{\mathbf{q}} -2 \Omega {\mathbf{q}} \cdot {\mathbf{P}}_0 + 
\Omega^2
P_{00} \nonumber\\
&=& \frac{2}{d}q^2 \sum_{\mathbf{k}}
  \frac{\Delta_{sc}^2}{E_{\mathbf{k}}^2} \left[
    \frac{1-2f(E_{\mathbf{k}})}{2E_{\mathbf{k}}} + f^\prime
    (E_{\mathbf{k}}) \right]\left[\varphi_{\mathbf{k}}^2 (\vec{\nabla}
    \epsilon_{\mathbf{k}})^2 - \frac{1}{4} (\vec{\nabla}
    \epsilon_{\mathbf{k}}^2)\cdot 
(\vec{\nabla}\varphi_{\mathbf{k}}^2)\right]
  \nonumber\\ 
&& -2\Omega^2 \sum_{\mathbf{k}} \left\{
    \frac{\Delta_{sc}^2\varphi_{\mathbf{k}}^2} {E_{\mathbf{k}}^2} 
\left[
      \frac{1-2f(E_{\mathbf{k}})}{2E_{\mathbf{k}}} + f^\prime
      (E_{\mathbf{k}})- f^\prime (E_{\mathbf{k}}) \frac{\Omega^2
        -({\mathbf{q}}\cdot \vec{\nabla}\epsilon_{\mathbf{k}})^2 - 
\Delta^2
        ({\mathbf{q}}\cdot \vec{\nabla} \varphi_{\mathbf{k}})^2} 
{\Omega^2 -
        ({\mathbf{q}}\cdot \vec{\nabla} E_{\mathbf{k}})^2}
    \right]\right.\nonumber\\ 
&&\left.
    +\frac{\Delta_{pg}^2}{4E_{\mathbf{k}}^2} f^\prime 
(E_{\mathbf{k}})
    \frac{({\mathbf{q}}\cdot \nabla\epsilon_{\mathbf{k}}^2)
      ({\mathbf{q}}\cdot \vec{\nabla} \varphi_{\mathbf{k}}^2) + 
\Delta^2
      ({\mathbf{q}}\cdot \vec{\nabla} \varphi_{\mathbf{k}}^2)^2} 
{\Omega^2 -
      ({\mathbf{q}}\cdot \vec{\nabla} E_{\mathbf{k}})^2}\right\}\,,
\label{q.P.q}
\end{eqnarray}
where $f(E)$ is the Fermi function.
Because Eq.~(\ref{q.P.q}) is ill-behaved for long wavelengths and low
frequencies, in order to calculate the AB mode velocity one needs to take
the appropriate limit $\Omega =c q \rightarrow 0$.  By contrast, the
calculation of the London penetration depth first requires to set $\Omega=0$
(static limit), and then $\textbf{q}\rightarrow 0$. The superfluid density
$n_s$ can be calculated from the coefficient of the $q^2$ term in Eq.~
(\ref{q.P.q}), (see, also Eq.~(\ref{Q00}) for $Q=0$).  Finally, the AB mode
``sound'' velocity $c=\Omega/q$, in the absence of the amplitude
renormalization, can be obtained by solving $q_\mu K_0^{\mu\nu}q_\nu =0$.

In the absence of the pseudogap (i.e., when $\Delta_{sc}=\Delta$) the last
term inside $\{\ldots\}$ in Eq.~(\ref{q.P.q}) drops out, and the resulting
analytical expression reduces to the standard BCS result
\cite{aronov76-498}, which at $T=0$ has the relatively simple form

\begin{equation}
  q_\mu K_0^{\mu\nu}q_\nu = \frac{q^2}{d} \sum_{\mathbf{k}}
  \frac{\Delta_{sc}^2}{E_{\mathbf{k}}^3} \left[\varphi_{\mathbf{k}}^2
    (\vec{\nabla} \epsilon_{\mathbf{k}})^2 - \frac{1}{4} (\vec{\nabla}
    \epsilon_{\mathbf{k}}^2)\cdot
    (\vec{\nabla}\varphi_{\mathbf{k}}^2)\right] -\Omega^2
  \sum_{\mathbf{k}} \frac{\Delta_{sc}^2\varphi_{\mathbf{k}}^2}
  {E_{\mathbf{k}}^3}\,.
\label{qPqatT=0}
\end{equation}
At finite $T$, the AB mode becomes damped, and the real and imaginary parts
of the sound velocity have to be calculated numerically.  Although, the
algebra is somewhat complicated, it can be shown that the AB mode satisfies
$ c \rightarrow 0 $ as $T \rightarrow T_c $, as expected.

\section{Numerical Results: Zero and Finite Temperatures}
\label{sec:CM_num_res}

In this section we summarize numerical results obtained for the AB mode
velocity $c$ associated with the electromagnetic response kernel, as
obtained by solving Eqs.~(\ref{eq:em19}) and (\ref{eq:em20}). We also
briefly discuss the behavior for the $T=0$ phase mode velocity $v_{\phi}$
which results from the coupling to density fluctuations, as well [see
Eq.~(\ref{eq:em24})]. The former, which has physical implications for the
behavior of the dielectric constant \cite{prange63-2495,zha95-6602}, is the
more straightforward to compute, because it does not require any new
approximations associated with the effective interactions $V$ in the
particle-hole channel.  The analysis of this section provides information
about the nature of the ``quasi-ideal'' Bose gas limit, which we address via
plots of the infinite $g$ asymptote of the AB mode, called $c_{\infty}$.
It also helps to clarify how pair fluctuations contribute, at finite
temperatures, to the collective mode dispersion.  Our $T=0$ calculations are
based on the Leggett ground state which corresponds to that of the pairing
approximation as well. At finite $T$, we numerically evaluate the AB sound
dispersion from Eq.~(\ref{q.P.q}), obtained within the framework of the
pairing approximation.

In Fig.~4 we plot the zero temperature value of $c$ as a function of the
dimensionless coupling strength $g/g_c$, where $g_c=-4\pi/m k_0$ is the
critical value of the coupling above which bound pairs are formed in vacuum.
Here we consider a 3D jellium model at three different electron densities,
which are parameterized via $k_0/k_F$. The most interesting feature of these
and related curves is shown in the inset where we plot the asymptotic limit
for each value of density or, equivalently, $k_o$.  This numerically
obtained asymptote reflects the \textit{effective} residual boson-boson
interactions in the ``quasi-ideal'' Bose gas limit, and is close to the
value calculated in Ref.~\onlinecite{engelbrecht97-15153} whose functional
dependence is given by $c_\infty/v_F\propto\sqrt{k_F/k_0}$ or, equivalently,
$c_\infty\propto\sqrt{n/k_0}$.  Interpreting the physics as if the system
were a true interacting Bose system, one would obtain an effective
interaction $U(0)\approx 3\pi^2/mk_0$, \textit{independent of $g$ in the
  strong coupling limit}.  As expected, these inter-boson interactions come
exclusively from the underlying fermion character of the system, and can be
associated with the repulsion between the fermions due to the Pauli
principle.  All of this is seen most directly \cite{haussmann93-291} by
noting that the behavior displayed in the inset can be interpreted in terms
of the effective scattering length of the bosons $a_B$ which is found to be
twice that of the fermions $a_F$ in the strong coupling limit.  Effects
associated with the coupling constant $g$ are, thus, entirely incorporated
into making bosons out of a fermion pair, and are otherwise invisible.

The same calculations are repeated in Fig.~5 for a tight binding lattice
bandstructure at $T=0$.  Figure~5(a) plots the sound velocity for different
densities $n$, as a function of the coupling constant; the behavior of the
large $g$ limit is shown in Fig.~5(b) as a function of density for a fixed
$g$.  Near half filling, where there is particle-hole symmetry, the
amplitude contributions are irrelevant and the large $g$ limit for $c$,
which follows from Eq.~(\ref{qPqatT=0}), is $c=\sqrt{2}t$, where $t$ is the
hopping integral.  At low $n$ the AB velocity varies as $\sqrt{n}$, which is
consistent with the results shown above for jellium.  In both cases the
behavior again reflects the underlying fermionic character, since it is to
be associated with a Pauli principle induced repulsion between bosons.
Unlike in the jellium case, where $c$ approaches a finite asymptote as $g$
increases, here $c$ vanishes asymptotically due to the increase of the pair
mass associated with lattice effects.\cite{NSR,chen99-7083} For
completeness, we also show, as an inset in Fig.~5(b), the behavior of
$v_{\phi}$, where we have used the RPA approximation to characterize the
parameter $V$ in the particle-hole channel. This approximation is in the
spirit of previous work by Belkhir and Randeria\cite{belkhir92-5087},
although it cannot be readily motivated at sufficiently large $g$.

Finally, in Fig.~6 we plot the temperature dependence of the AB mode
velocity (both real and imaginary parts), for moderately strong coupling
(solid lines) and the BCS limit (dashed lines). For the former, our curves
stop somewhat below $T_c$, since close to the critical temperature,
inaccuracies are introduced by our neglect of the temperature dependence of
the amplitude mode contributions to $c$. It should be noted that the
transition from a finite to zero value for the ``sound'' velocity appears to
be rather abrupt in the vicinity of $T_c$, the stronger the coupling. This
figure suggests that the AB mode velocity reflects the same transition
temperature $T_c$ as is computed via the excited pair propagator, or
T-matrix.  This represents an important self consistency check on the
present formalism.

\section{Conclusions}
\label{sec:conc}

This paper deals with the fairly complex issues of pair fluctuations,
collective modes and gauge invariance in a BCS Bose-Einstein crossover
scenario. A starting point for our approach is the Leggett ground state,
which imposes rather strong constraints on the nature of the physics of
fermions and composite bosons.  The fermion degrees of freedom are always
present through the self consistency conditions and can never be fully
integrated out.  Not only are these fermion pairs different from true
bosons, but they represent a very special type of composite boson.  They
behave rather specifically, in correspondence with the underlying structure
of the BCS state. Even at $T=0$ one can see from previous work on the
BCS-BEC crossover, that these constraints lead to a mix of ideal Bose gas
\cite{Leggett} and non-ideal \cite{belkhir92-5087} Bose liquid behavior.
This mirrors some of the effects of BCS theory, in which the system
undergoes a form of Bose condensation with a full condensate fraction $n_0 /
n = 1$, at $T=0$.  Nevertheless, a BCS superconductor has a sound-like
collective (phase) mode which is intimately associated with its
superconductivity.  Moreover, as is discussed in Sec.~V.A, it is instructive
to contrast the polarizability associated with the phase mode, called
$Q'_{22}(Q) $, with the pair susceptibility $\chi(Q)$.  These two modes
correspond to distinct dynamical branches, although both become soft at
$\mathbf{q}=0$ under the same conditions.  The softness of the former is
naturally associated with the Goldstone boson and the latter with a
vanishing chemical potential for pairs: $\mu_{pair} = 0$.

A central physics theme of this paper is reflected in Fig.~1. The crossover
problem introduces an important new parameter $\Delta_{pg}$ which
characterizes the excited states of a non-BCS superconductor. As a result,
there are three coupled equations to be satisfied for $\Delta_{pg}$,
$\Delta$ and $\mu$, at finite $T$, in contrast to zero temperature crossover
theories where there are only two.  This new parameter is a measure of the
difference between the excitation gap and the superconducting order
parameter. One can also arrive at a picture which is similar to Fig.~1
within the Coulomb-modulated phase fluctuation scenario\cite{emery95-434}.
What is different here is the ``tuning parameter'', which corresponds to the
coupling strength $g$ in the crossover picture, and the plasma frequency
$\omega_p$ in the phase fluctuation scenario.

One may summarize our results by asking the following series of questions,
which our paper raises and answers.

\textit{Is there any precedent for soft modes other than Goldstone bosons?}

Yes. In the present paper we find that a vanishing value for the pair
chemical potential $\mu_{pair}$ below $T_c$ leads to a soft mode 
corresponding to
incoherent, finite momentum pair excitations.  These excitations are
analogous to the ``particle'' excitations in the case of the neutral Bose
liquid.  Moreover, for the Bose liquid, this branch is also soft and always
different (except at zero wavevector) from the Goldstone boson.  However, in
the true Bose system the two branches have the same slope at $\textbf{q}=0$.

\textit{Should the T-matrix ${\cal T }$ be renormalized so as to yield the
  sound mode dispersion at small wave-vectors, as in a Bose liquid?} With
this renormalization ${\cal T}$ would then be similar to its Bose liquid
counterpart\cite{fetter} and in this way the pair excitation dispersion
$\Omega_{\bf q}$ at small wavevectors would be linear in ${\bf q}$.  We
answer this question by noting that the fermion degrees of freedom strongly
constrain ${\cal T }$ so that the composite boson system is different from
the Bose liquid.  As a consequence the proposed renormalization seems
problematical for two reasons. Adding in this collective mode effect
associated with the Bose liquid is equivalent to including boson-boson
interactions.  In the present composite boson case, these boson-boson
interactions derive from fermionic degrees of freedom, i.e., the Pauli
principle, which have already been accounted for in our calculations of
${\cal T }$. It is not clear why these boson-boson interactions should then
be included yet a second time.
Secondly, once there is a renormalization of ${\cal T }$, this will
change the ground state gap equation and corresponding constraint on the
fermionic chemical potential $\mu$, associated with the changed
fermionic self energy $\Sigma$.  Moreover, this renormalization will,
presumably, introduce an unphysical incomplete condensation in the
ground state, like its counterpart in the Bose liquid.  In summary, the
composite boson system is considerably different from the true boson
system, because the underlying fermionic constraints (through $\Delta$,
$\mu$) can never be ignored.

\textit{What about Coulomb renormalizations of the pair propagator?}

If Coulomb interactions were included, it would follow, if the analogy were
appropriate, that the pair fluctuation mode would then be gapped.  This
would again compromise the self consistent conditions which must be
satisfied from the fermionic perspective (i.e., $\Delta$ and $\mu$) in the
Leggett ground state.  Indeed, within the BCS formalism (as well as in the
Leggett ground state) long range Coulomb interactions do not enter in an
important way to change the gap equation structure, but rather they
principally affect the collective modes.  It is for this reason that we
argued earlier that Coulomb effects are presumed to be already included in
the pairing interaction.  It should be recalled that
at large $g$, we have seen that
essentially all signs of the two body ($g$ dependent) fermion-fermion
interaction are absent in the effective boson-boson interaction.

\textit{Does this paper in any way change the way we think about BCS
  theory?} Absolutely not. BCS theory appears as a special case in the weak
coupling $g$, particle-hole symmetric limit of our more general approach.
When we study pair excitations in this limit, we find they are greatly
damped at all wavevectors $q$, and it makes no sense to talk about them. In
this way $\Delta_{pg}(T_c)/\Delta_{sc}(0)$ is vanishingly small in the BCS
limit.

\textit{To what extent are the results of this paper limited by the T matrix
  approximation?} The T-matrix approximation seems to be intimately
connected to the physics of the crossover scenario. This scheme represents,
in some sense, a truncation of the interactions at a pair-wise level. This
truncation, which appears to generate a quasi-ideal Bose gas character to
the composite boson system, mirrors the behavior of the well established
ground state. This approach might not be suitable for other composite boson
scenarios, which do not evolve directly from the BCS phase. Nevertheless,
these crossover schemes provide a useful way of learning about composite
boson systems in general.  Moreover, they provide valuable insights about
how to extend BCS theory slightly, without abandoning it altogether, and in
this way to address a large class of short coherence length, but otherwise
conventional, superconductors.

\acknowledgements

We are very grateful to Gene Mazenko, Andrei Varlamov and Paul Wiegmann for
useful discussions and to Igor Kulik for helpful communications.  This work
was supported by grants from the National Science Foundation through the
Science and Technology Center for Superconductivity under Grant
No.~DMR~91-20000 and through the MRSEC under Grant No.~DMR~9808595, 5-43030.

\appendix

\section{Evaluation of the Vertex Corrections}
\label{ap:1}

In this appendix we demonstrate an explicit cancellation between the
Maki-Thompson (MT) and Aslamazov-Larkin (AL) diagrams of Fig.~3.  In
this way we prove that the contribution to the vertex correction $\delta
\Lambda$ from the superconducting order parameter is given by the
Maki-Thompson diagram, and the pseudogap contribution $\delta
\Lambda_{pg}$ comes from Maki-Thompson (MT) and Aslamazov-Larkin (AL)
diagrams.  It is easy to demonstrate a cancellation between the MT
diagram and the AL diagrams, which will greatly simplify the
calculations. In general, we have
\begin{equation}
  \delta \Lambda_{pg}^\mu (K,K-Q)q_\mu = -(MT)_{pg} + \sum_P
  {\cal T}_{pg}(P)G_0 (P-K) \frac{\partial
    \varphi^2_{{\mathbf{k}}-{\mathbf{p}}/2-{\mathbf{q}}/2}} 
   {\partial {\bf k}}\cdot {\bf q}\,,
\label{cancellation}
\end{equation}
where $(MT)_{pg}$ refers to the MT diagram contribution, and
${\cal T}_{pg}(Q\neq 0)$ is the $T$ matrix or pair propagator.


To prove this cancellation, we notice that the vertex corrections in
the (four-)current-current correlation functions can be
obtained from proper vertex insertions in the single particle Green's
functions in the self-energy diagram.  In the pairing
approximation ( $G_oG$ scheme) we have
\begin{equation}
\Sigma_{pg}(K)=\sum_L {\cal T}_{pg}(K+L)
G_0 (L) \varphi^2_{(K-L)/2},
\label{Eq:self-energy}
\end{equation}
where $L$ is the four-momentum of the fermion loop, this procedure leads
to one Maki-Thompson diagram and two Aslamazov-Larkin diagrams. 

Obviously, $L$ in the Eq.~(\ref{Eq:self-energy}) is a dummy variable so
that its variation does not change $\Sigma(K)$, namely,
\begin{eqnarray}
  0&=&\sum_L \left[ {\cal T}_{pg}((K+L+\Delta L)G_0(L+\Delta L)
    \varphi^2_{(K-L-\Delta L)/2} - {\cal T}_{pg}((K+L)G_0(L)
    \varphi_{(K-L)/2}^2\right]\nonumber\\ 
  &=&\sum_L \left\{\left[{\cal T}_{pg}(K+L+\Delta L)-{\cal T}_{pg}(K+L)\right] G_0(L+\Delta
    L) \varphi^2_{(K-L-\Delta L)/2}\right.\nonumber\\ 
  && \left.
    +{\cal T}_{pg}(K+L) \left[G_0(L+\Delta L) -G_0(L)\right] \varphi^2_{(K-L-\Delta
      L)/2} + {\cal T}_{pg}(K+L) G_0(L) \left[ \varphi^2_{(K-L-\Delta L)/2} -
      \varphi^2_{(K-L)/2} \right]\right\}
\label{Eq:diff_Sigma}
\end{eqnarray}
Using $G(K)G^{-1}(K)=1$, we obtain 
\begin{mathletters}
\label{Eq:vertex-insertion}
\begin{eqnarray}
G(K+\Delta K)-G(K) &=& -G(K)[G^{-1}(K+\Delta K) - G^{-1} (K)] G(K+\Delta
K)\nonumber\\
&=&-G(K)\Lambda_\mu (K+\Delta K, K) G(K) \Delta K^\mu \;,
\end{eqnarray}
where $G^{-1}(K+\Delta K) - G^{-1} (K)\approx \Lambda_\mu (K+\Delta K, K)
\Delta K^\mu$ is the full vertex. Similarly, we have
\begin{eqnarray}
  G_0(K+\Delta K)-G_0(K) &=& -G_0(K)[G_0^{-1}(K+\Delta K) - G_0^{-1}
  (K)] G_0(K+\Delta K)\nonumber\\ 
  &=&-G_0(K)\lambda_\mu (K+\Delta K, K)G_0(K) \Delta K^\mu \;,
\end{eqnarray}
\end{mathletters}
where $G_0^{-1}(K+\Delta K) - G_0^{-1} (K)\approx \lambda_\mu (K+\Delta
K, K) \Delta K^\mu$ is the bare vertex, and $\lambda^\mu (K+\Delta K, K)=(1,
\vec \nabla_{\mathbf{k}} \epsilon_{{\mathbf{k}+\Delta {\mathbf{k}/2}}})$.

Equations (\ref{Eq:vertex-insertion}) correspond to the vertex insertions
diagrammatically along the full and bare Green's functions, respectively.

Using ${\cal T}_{pg}(K+L)=g/[1+g\chi (K+L)]$, we obtain
\begin{equation}
{\cal T}_{pg}(K+L+\Delta L)-{\cal T}_{pg}(K+L)= -{\cal T}_{pg}(K+L+\Delta L)[\chi(K+L+\Delta L)-\chi(K+L)]{\cal T}_{pg}(K+L)
\label{Eq:diff_t}
\end{equation}
Writing $\chi(K+L)=\sum_{L^\prime} G(L^\prime)G_0(K+L-L^\prime)
\varphi^2_{L^\prime-(K+L)/2}$, we have
\begin{eqnarray}
\chi(K+L+\Delta L)-\chi(K+L)&=& \sum_{L^\prime} G(L^\prime)
\left\{\left[G_0(K+L-L^\prime+\Delta L)-G_0(K+L-L^\prime)\right]
  \varphi^2_{L^\prime-(K+L+\Delta L)/2}\right.\nonumber\\
&&\left. + G_0(K+L-L^\prime)\left[
    \varphi^2_{L^\prime-(K+L+\Delta L)/2}
    -\varphi^2_{L^\prime-(K+L)/2}\right] \right\}
\label{Eq:diff_chi1}
\end{eqnarray}
On the other hand, writing $\chi(K+L)=\sum_{L^\prime}
G(K+L-L^\prime)G_0(L^\prime) \varphi^2_{(K+L)/2-L^\prime}$, we get
\begin{eqnarray}
\chi(K+L+\Delta L)-\chi(K+L)&=& \sum_{L^\prime} 
\left\{\left[G(K+L-L^\prime+\Delta L)-G(K+L-L^\prime)\right]G_0(L^\prime)
  \varphi^2_{(K+L+\Delta L)/2-L^\prime}\right.\nonumber\\
&&\left. + G(L^\prime)G_0(K+L-L^\prime)\left[
    \varphi^2_{L^\prime-(K+L-\Delta L)/2}
    -\varphi^2_{L^\prime-(K+L)/2}\right] \right\}
\label{Eq:diff_chi2}
\end{eqnarray}
Combining Eq.~(\ref{Eq:diff_chi1}) and Eq.~(\ref{Eq:diff_chi2}), we
obtain to the first order of $\Delta L$
\begin{eqnarray}
\chi(K+L+\Delta L)-\chi(K+L)&=& \frac{1}{2} \sum_{L^\prime} 
\left\{ G(L^\prime)\left[G_0(K+L-L^\prime+\Delta
    L)-G_0(K+L-L^\prime)\right]\varphi^2_{L^\prime-(K+L+\Delta L)/2}
\right.\nonumber\\ 
&&+\left.
\left[G(K+L-L^\prime+\Delta L)-G(K+L-L^\prime)\right]G_0(L^\prime)
  \varphi^2_{L^\prime-(K+L+\Delta L)/2}\right\}\;,
\label{Eq:diff_chi}
\end{eqnarray}
where we have assumed in general $\varphi^2_K =\varphi^2_{-K}$.
Substituting Eq.~(\ref{Eq:diff_chi}) and  Eq.~(\ref{Eq:diff_t})
into Eq.~(\ref{Eq:diff_Sigma}), we obtain
\begin{eqnarray}
0&=&-\frac{1}{2} \sum_{L L^\prime} {\cal T}_{pg}(K+L+\Delta L){\cal T}_{pg}(K+L) \left\{
  G(L^\prime) \left[ G_0(K+L-L^\prime+\Delta
    L)-G_0(K+L-L^\prime)\right]\varphi^2_{L^\prime-(K+L+\Delta
  L)/2}\right. \nonumber\\
&&\left. +\left[G(K+L-L^\prime+\Delta L)-G(K+L-L^\prime)\right]G_0(L^\prime)
  \varphi^2_{L^\prime-(K+L+\Delta L)/2}\right\} G_0 (L+\Delta L)
  \varphi^2_{(K-L-\Delta L)/2}\nonumber\\
&& + \sum_L  {\cal T}_{pg}(K+L) \left[ G_0 (L+\Delta L) -G_0(L)\right]
  \varphi^2_{(K-L-\Delta L)/2} \nonumber\\
&& + \sum_L {\cal T}_{pg}(K+L)G_0(L) \left[\varphi^2_{(K-L-\Delta L)/2} -
  \varphi^2_{(K-L)/2}\right]
\label{Eq:diff_Sigma2}
 \end{eqnarray}
 Comparing this with the analytical expressions corresponding to the
 diagrams in Fig.~3, it is easy to identify the first two terms
 as the two AL diagrams (which we denote by $AL_1$ and $AL_2$) and the
 third one with the MT diagram for the pseudogap vertex corrections.
 Therefore,
\begin{equation}
  \frac{1}{2}\left[(AL_1) + (AL_2)\right] + (MT)_{pg} + \sum_L
  {\cal T}_{pg}(K+L)G_0(L)\left[\varphi^2_{(K-L-\Delta L)/2} -
    \varphi^2_{(K-L)/2} \right] =0
\end{equation}
Finally, we have
\begin{eqnarray}
  \delta \Lambda_{pg}^\mu(K, K-\Delta L) \Delta L_\mu &=& (AL_1) +(AL_2) +
  (MT)_{pg}\nonumber\\ &=&-(MT)_{pg}-2\sum_L {\cal T}_{pg}(K+L)G_0(L)\frac{\partial
    \varphi^2_{(K-L-\Delta L)/2}} {\partial L}\cdot \Delta L
\end{eqnarray}
Changing variables $K+L\rightarrow P, \Delta L \rightarrow Q$ leads to
Eq.~(\ref{cancellation}).

The two contributions which enter Eq.~(\ref{Vertex}) result from
adding the superconducting gap and pseudogap terms which are given
respectively by
\begin{mathletters}
\begin{equation}
\delta \Lambda_{sc} (K, K-Q) = -\Delta_{sc}^2\varphi_{\mathbf{k}} 
\varphi_{\mathbf{k-q}}G_0(-K)G_0(Q-K)\lambda(Q-K,-K)\,,
\end{equation}
and
\begin{eqnarray}
  \delta \Lambda_{pg}^\mu (K, K-Q)&=&-\sum_P
  {\cal T}_{pg}(P)\varphi_{\mathbf{k-p/2}} \varphi_{\mathbf{k-q-p/2}}
  G_0(P-K)G_0(P+Q-K) \lambda^\mu (P+Q-K,P-K) \nonumber\\ & & + 
\sum_P
  {\cal T}_{pg}(P)G_0 (P-K) \frac{\partial
    \varphi^2_{{\mathbf{k}}-{\mathbf{p}}/2-{\mathbf{q}}/2}} 
{\partial
    k_\mu}\,,
\end{eqnarray}
\end{mathletters}

\section{Full Expressions for the correlation functions $\tensor{P}$,
  $\mathbf{P}_0$ and $P_{00}$}

\label{ap:2}

It is useful here to write down the component contributions to the
different correlation functions in the electromagnetic response.  After
adding the superconducting and pseudogap contributions one finds for the
current-current correlation function

\begin{eqnarray}
  \tensor{\frac{n}{m}}+\tensor{\bf P} & = & 2\sum_{\mathbf{k}}
  \frac{\Delta_{sc}^2}{E_{\mathbf{k}}^2} \left[
    \frac{1-2f(E_{\mathbf{k}})}{2E_{\mathbf{k}}} + f^\prime
    (E_{\mathbf{k}}) \right] \left[\varphi_{\mathbf{k}}^2 
(\vec{\nabla}
    \epsilon_{\mathbf{k}})(\vec{\nabla} \epsilon_{\mathbf{k}}) -
    \frac{1}{4} (\vec{\nabla} \epsilon_{\mathbf{k}}^2)
    (\vec{\nabla}\varphi_{\mathbf{k}}^2)\right] 
\nonumber\\ 
&&-2\sum_{\mathbf{k}} f^\prime (E_{\mathbf{k}}) 
\frac{\Omega^2}{\Omega^2
  - ({\mathbf{q}}\cdot \vec{\nabla} E_{\mathbf{k}})^2} (\vec{\nabla}
\epsilon_{\mathbf{k}}) (\vec{\nabla}
\epsilon_{\mathbf{k}}) \nonumber\\ 
&&+ \sum_{\mathbf{k}} \frac{\Delta_{pg}^2}{E_{\mathbf{k}}^2} 
f^\prime
(E_{\mathbf{k}}) \frac{\Omega^2}{\Omega^2 - ({\mathbf{q}}\cdot 
\vec{\nabla}
  E_{\mathbf{k}})^2} \left[\varphi_{\mathbf{k}}^2 (\vec{\nabla}
  \epsilon_{\mathbf{k}})(\vec{\nabla} \epsilon_{\mathbf{k}}) -
  \frac{1}{4} (\vec{\nabla} \epsilon_{\mathbf{k}}^2)
  (\vec{\nabla}\varphi_{\mathbf{k}}^2)\right] \,,
\label{Q00}
\end{eqnarray}
and for the current-density correlation function
\begin{equation}
  {\mathbf{P}}_0=-2\Omega \sum_{\mathbf{k}} 
\frac{\epsilon_{\mathbf{k}}}
  {E_{\mathbf{k}}} f^\prime (E_{\mathbf{k}}) \frac{{\mathbf{q}}\cdot
    \vec{\nabla} E_{\mathbf{k}}} {\Omega^2 - ({\mathbf{q}}\cdot 
\vec{\nabla}
    E_{\mathbf{k}})^2} \vec{\nabla} \epsilon_{\mathbf{k}} \,,
\label{Q03}
\end{equation}
and finally for the density-density correlation function
\begin{equation}
  P_{00}=-2\sum_{\mathbf{k}}
  \frac{\Delta_{sc}^2\varphi_{\mathbf{k}}^2}{E_{\mathbf{k}}^2} \left[
    \frac{1-2f(E_{\mathbf{k}})}{2E_{\mathbf{k}}} + f^\prime
    (E_{\mathbf{k}}) \right] + 2\sum_{\mathbf{k}} f^\prime
  (E_{\mathbf{k}}) \frac{\Omega^2 
\Delta_{sc}^2\varphi_{\mathbf{k}}^2 -
    E_{\mathbf{k}}^2 ({\mathbf{q}}\cdot \vec{\nabla} 
E_{\mathbf{k}})^2}
    {E_{\mathbf{k}})^2 \left[ \Omega^2 - ({\mathbf{q}}\cdot 
\vec{\nabla}
        E_{\mathbf{k}})^2\right]} \,.
\label{Q33}
\end{equation}
In deriving the first of these we have integrated by parts
to evaluate
\begin{eqnarray}
  \frac{n}{m} &=& 2\sum_K \frac{\partial^2 \epsilon_{\mathbf{k}}
    }{\partial {\mathbf{k}} \partial
  {\mathbf{k}}} G(K) 
= -2  \sum_{\mathbf{k}} G^2(K) (\vec{\nabla}
\epsilon_{\mathbf{k}})\cdot \left[ \vec{\nabla}
\epsilon_{\mathbf{k}} + \vec{\nabla} \Sigma(K)\right]\nonumber\\
&=& 2\sum_{\mathbf{k}} \frac{\Delta^2}{E_{\mathbf{k}}^2} \left[
  \frac{1-2f(E_{\mathbf{k}})}{2E_{\mathbf{k}}} + f^\prime
  (E_{\mathbf{k}}) \right] \left[\varphi_{\mathbf{k}}^2 (\vec{\nabla}
  \epsilon_{\mathbf{k}})\cdot(\vec{\nabla} \epsilon_{\mathbf{k}}) -
  \frac{1}{4} (\vec{\nabla} \epsilon_{\mathbf{k}}^2)\cdot
  (\vec{\nabla}\varphi_{\mathbf{k}}^2)\right] -2\sum_{\mathbf{k}}
f^\prime (E_{\mathbf{k}})(\vec{\nabla} \epsilon_{\mathbf{k}})\cdot
(\vec{\nabla} \epsilon_{\mathbf{k}}) \,.
\label{n/m}
\end{eqnarray}
These expressions are then used to evaluate Eq.~(\ref{q.P.q}) in the 
text.


\begin{figure}
\centerline{\epsfxsize=3.5in\epsffile{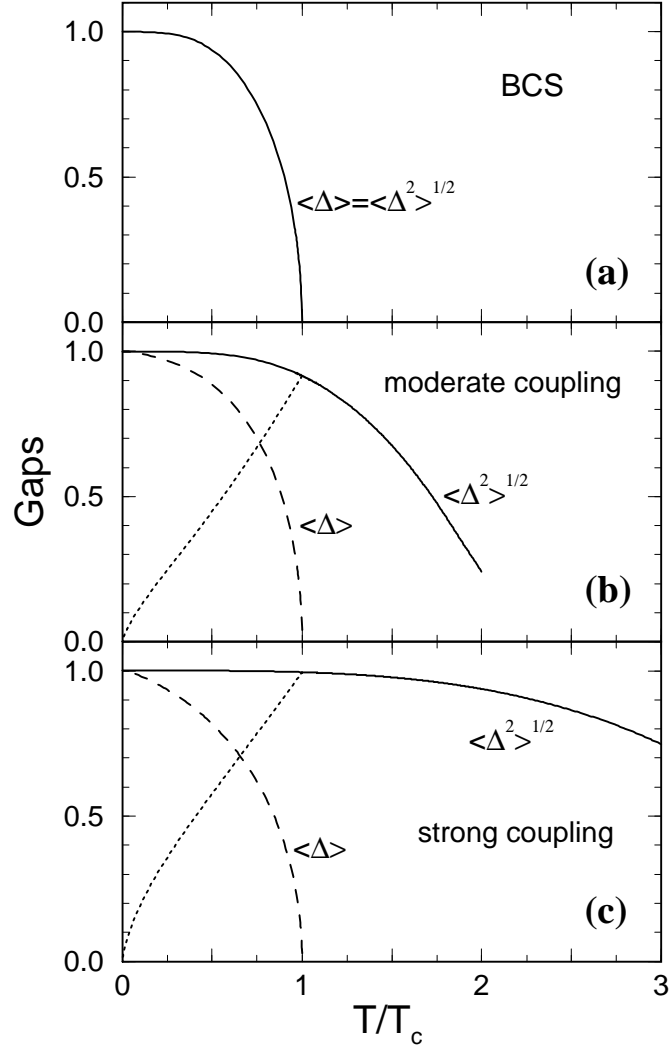}}
\caption{Temperature dependence of the excitation gap
  $\langle\Delta^2\rangle^{1/2}$ and the order parameter
  $\langle\Delta\rangle$ (normalized at $T=0$) for (a) weak coupling BCS,
  (b) moderate coupling, and (c) strong coupling. The dotted lines represent
  the difference of these two energy scales, corresponding to the pseudogap
  parameter $\Delta_{pg}$.  A strong pseudogap develops as the coupling
  strength increases.}
\label{Fig1}
\end{figure}

\begin{figure}
\centerline{\epsfxsize=3.5in\epsffile{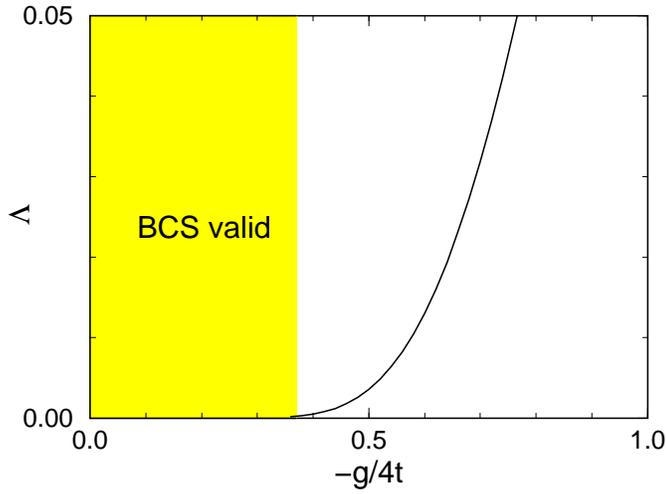}}
\caption{Dependence of the cutoff momentum $\Lambda$, (where the pair
  excitation spectrum crosses into the particle-particle continuum), on the
  coupling strength on a quasi-2D lattice.  Here $4t$ is the half bandwidth.
  Because of strong damping of the pair excitations, they are irrelevant at
  low $g$ and BCS theory is valid in the shaded region. Outside this regime,
  pair excitations become important.}
\label{Fig2}
\end{figure}

\begin{figure}[htbp]
\centerline{\epsfxsize=3.5in\epsffile{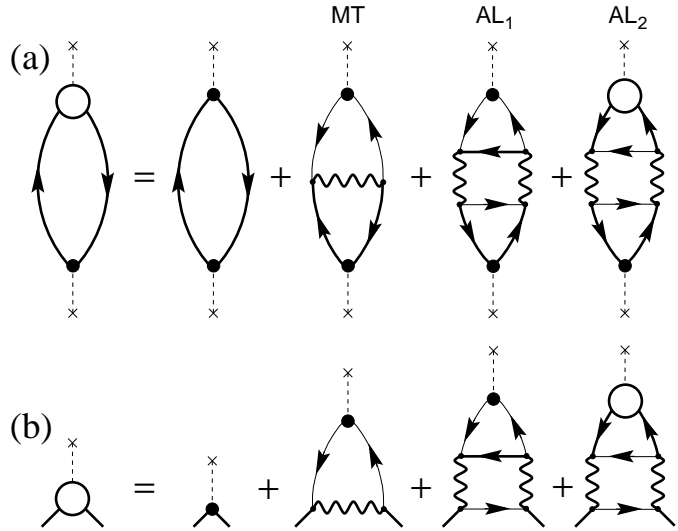}}
\medskip
  \caption{Diagramatic representation of (a) the polarization bubble, 
    and (b) the vertex function used to compute the electrodynamic
    response functions. Here the wavy lines represent ${\cal T }$ and it
    should be noted that the thin and thick lines correspond to $G_o$
    and $G$ respectively. The total vertex correction is given by the
    sum of the Maki Thompson (MT) and two Aslamazov-Larkin (AL$_1$ and
    AL$_2$) diagrams.}
  \label{fig:diagrams}
\end{figure}

\begin{figure}
\centerline{\epsfxsize=3.5in\epsffile{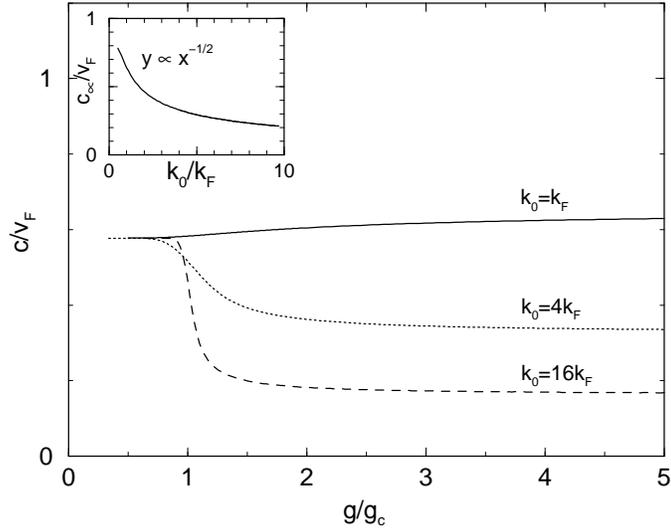}}
\caption{AB mode velocity $c/v_F$ as a function of the coupling strength (main
  figure) for various densities characterized by $k_0/k_F$ in 3D
  jellium. Plotted in the inset is the large $g$ asymptote
  $c_{\infty}/v_F$, versus $k_0/k_F$, which varies as $(k_F/k_0)^{1/2}$,
  as expected.  }
\label{Fig4}
\end{figure}

\begin{figure}
\centerline{\epsfxsize=3.5in\epsffile{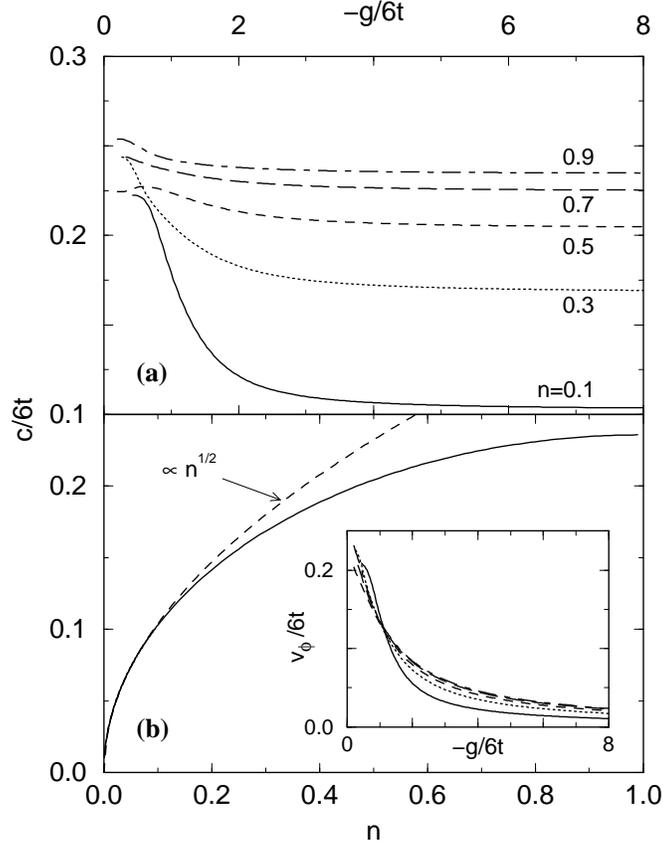}}
\caption{(a) Normalized
  AB mode velocity, $c/6ta$, on a 3D lattice with an $s$-wave pairing
  interaction for various densities as a function of $g$, and (b) the
  large $g$ limit for $c/6ta$ as a function of density $n$ for fixed
  $-g/6t=20$. (Here $6t$ is the half bandwidth). The dashed line in (b)
  shows a fit to the expected low density dependence $n^{1/2}$. Plotted
  in the inset is the velocity of the phase and density coupled
  collective mode $v_{\phi}/ 6ta$ with the particle-hole channel treated
  at the RPA level, for the same $n$ as in (a).  }
\label{Fig5}
\end{figure}

\begin{figure}
\centerline{\epsfxsize=3.5in\epsffile{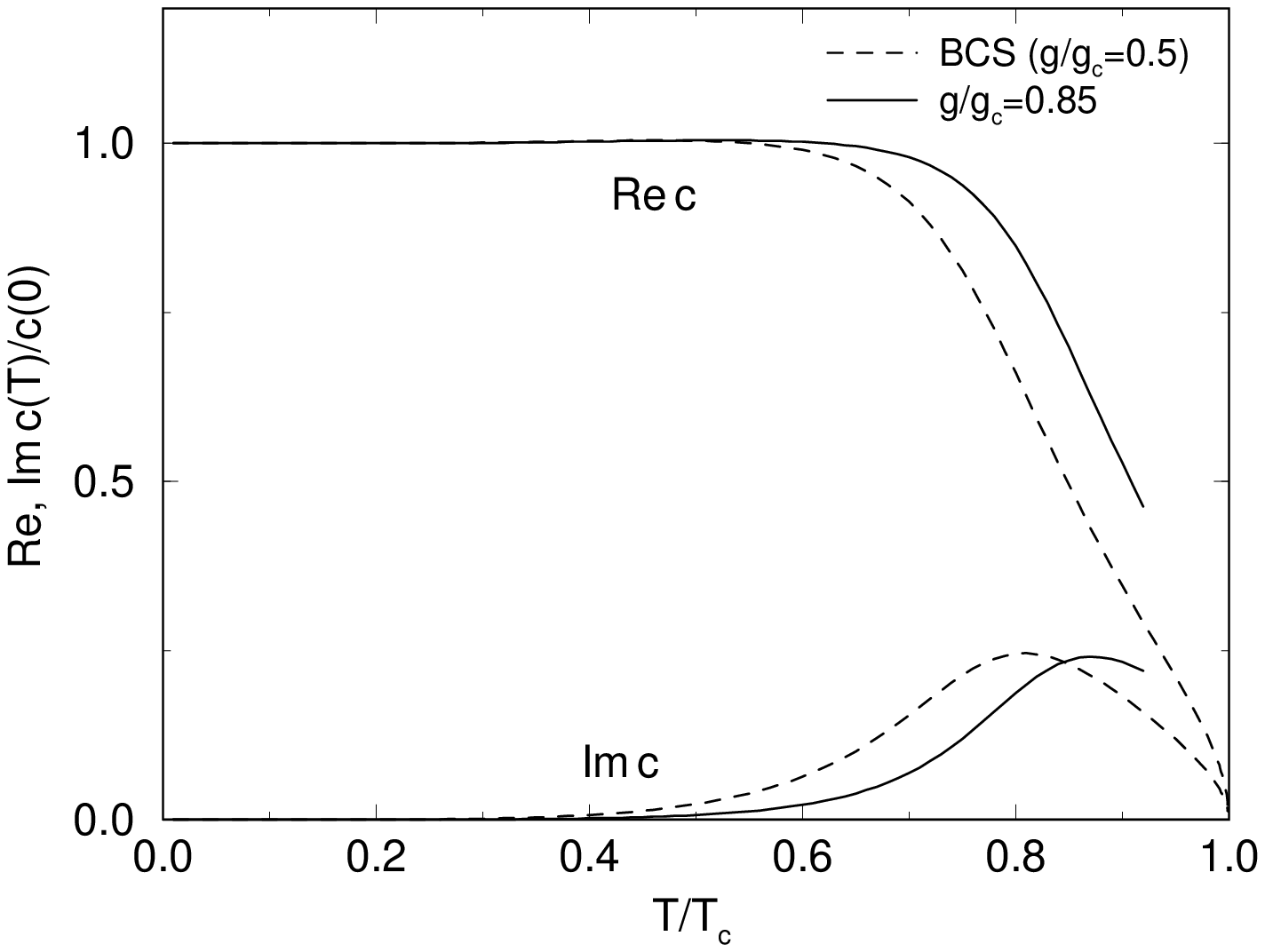}}
\caption{Temperature dependence of the real (Re$\,c$) and imaginary
  (Im$\,c$) parts of the AB mode velocity for moderate coupling
  (solid lines) and weak coupling BCS (dashed lines) in  3D
  jellium with $k_0=4k_F$. The mode is highly damped as $T_c$ is
  approached. Because of inaccuracy due to neglected T-
dependent amplitude effects,
the solid curves are cut off slightly below $T_c$.}
\label{Fig6}
\end{figure}


\begin{references}
\bibitem{AboveTcExtrapolation}
Above $T_c$, the gap equation discussed below in the text,
  Eq.~(\ref{Thouless-crit2}), no longer holds. Instead, the LHS should be
  equated with a $T$ dependent quantity, $\tau$. Slightly above $T_c$, this
  quantity is small ($\propto (T-T_c)$), therefore, we may still use the
  appoximation given by Eq.~(\ref{eq:sigma-pg}). However, the pair excitation
  spectrum will also acquire a $T$ dependent gap. In consequence, we have a
  modified set of equations Eqs.~(\ref{eq:5}) to solve for $\mu$,
  $\Delta_{pg}$, and $\tau$ above $T_c$. This extrapolation becomes worse as
  $T$ increases.
\end{references}
\end{document}